\documentclass[12pt,english]{article}
\pdfoutput=1
\usepackage[T1]{fontenc}
\usepackage[utf8]{inputenc}
\usepackage{geometry}
\usepackage{babel}

\usepackage{booktabs}
\usepackage{array}
\usepackage{longtable}
\usepackage{float}
\usepackage{amsmath}
\usepackage{graphicx}
\usepackage{amssymb}
\usepackage{esint}
\usepackage[unicode=true, pdfusetitle,
 bookmarks=true,bookmarksnumbered=false,bookmarksopen=false,
 breaklinks=false,pdfborder={0 0 0},backref=false,colorlinks=false]
 {hyperref}

\newcommand{\be}{\begin{equation}}
\newcommand{\ee}{\end{equation}}

\begin{document}

\title{Testing gaussianity, homogeneity and isotropy with the cosmic
microwave background}

\author{\normalsize L. Raul Abramo${}^1$ and Thiago S. Pereira${}^2$}

\footnotetext[1] {Instituto de F\'isica, Universidade de S\~ao Paulo, CP
66318, 05315-970, S\~ao Paulo, Brazil. E-mail: abramo@fma.if.usp.br}
\footnotetext[2] {Instituto de F\'isica Te\'orica, Universidade Estadual
Paulista, CP 70532-2, 01156-970, S\~ao Paulo, Brazil. E-mail:
thiago@ift.unesp.br}

\maketitle
\begin{abstract}
We review the basic hypotheses which motivate the statistical framework
used to analyze the cosmic microwave background, and how that framework can
be enlarged as we relax those hypotheses. In particular, we try to separate
as much as possible the questions of gaussianity, homogeneity and isotropy
from each other. We focus both on isotropic estimators of non-gaussianity as 
well as statistically anisotropic estimators of gaussianity, giving particular 
emphasis on their signatures and the enhanced ``cosmic variances'' that become 
increasingly important as our putative Universe becomes less symmetric. After reviewing
the formalism behind some simple model-independent tests, we discuss how these tests 
can be applied to CMB data when searching for large scale ``anomalies''.
\end{abstract}

\tableofcontents{}

\newpage{}

\section{Introduction}

According to our current understanding of the Universe, the morphology of
the cosmic microwave background (CMB) temperature field, as well as all
cosmological structures that are now visible, like galaxies,
clusters of galaxies and the whole web of large-scale structure, are
probably the descendants of quantum process that took place some $10^{-35}$
seconds after the Big Bang. In the standard lore, the machinery responsible
for these processes is termed cosmic inflation and, in general 
terms, what it means is that microscopic quantum fluctuations pervading the
primordial Universe are stretched to what correspond, today, to
cosmological scales (see \cite{Mukhanov:05,Weinberg:2008zzc,Peter:2009zzc}
for comprehensive introductions to inflation.) These primordial
perturbations serve as initial conditions for the process of structure
formation, which enhance these initial perturbations through gravitational
instability. The subsequent (classical) evolution of these instabilities 
preserves the main statistical features of these fluctuations that were
inherited from their inflationary origin -- provided, of course, that we
restrain ourselves to linear perturbation theory. 

However, given that matter has a natural tendency to cluster, and this
inevitably leads to non-linearities (not to mention the sorts of
complications that come with baryonic physics), the structures which 
are visible today are far from ideal probes of those statistical
properties. CMB photons, on the other hand, to an excellent
approximation experience free streaming since the time of decoupling ($z
\approx 1100$), and are therefore exempt from these non-linearities
(except, of course, for secondary anisotropies such as the Rees-Sciama
effect or the Sunyaev-Zel'dovich effect), which implies that they
constitute an ideal window to the physics of the early Universe -- see,
e.g., \cite{Hu:1997mn,Hu:2002aa,Dodelson:03}. 
In fact, we can determine the primary CMB anisotropies as well as most of
the secondary anisotropies on large scales, such as the Integrated
Sachs-Wolfe effect, completely in terms of the initial conditions by means
of a {\it linear kernel}:
\begin{equation}
\label{sources}
\Theta(\hat{n}) \equiv
\frac{\Delta T(\hat{n};\eta_0)}{T(\eta_0)}
= \int d^3 x' \int_0^{\eta_0} d\eta' \; \sum_i
K_i(\vec{x}\, ',\eta '; \hat{n}) {\cal{S}}^i (\vec{x} \, ',\eta') \; ,
\end{equation}
where $\eta'$ is conformal time, and ${\cal{S}}^i$ denote the initial
conditions of all matter and metric fields (as well as their time
derivatives, if the initial conditions are non-adiabatic.) Here $K_i$
is a linear kernel, or a retarded Green's function, that propagates the
radiation field to the time and place of its detection, here on
Earth. Since that kernel is insensitive to the statistical nature of the
initial conditions (which can be thought of as constants which multiply the
source terms), those properties are precisely transferred to the CMB
temperature field $\Theta$.

The statistical properties of the primordial fluctuations are, to lowest
order in perturbation theory, quite simple: because the 
quantum fluctuations that get stretched and enhanced by inflation
are basically harmonic oscillators in their ground state, the distribution
of those fluctuations is Gaussian, with each mode an independent
random variable. The Fourier modes of these fluctuations are 
characterized by random phases (corresponding to the random initial
values of the oscillators), with zero mean, and variances which are 
given simply by the field mass and the mode's wavenumber 
$k=2\pi/\lambda$. The presence of higher-order interactions (which exist
even for free fields, because of gravity) changes this simple picture, 
introducing higher-order correlations which destroy gaussianity
-- even in the simplest scenario of inflation
\cite{Maldacena:2002vr,Weinberg:2005vy,Weinberg:2006ac}.
However, since these interactions are typically suppressed by 
powers of the factor $GH^2\simeq10^{-12}$, where $G$ is Newton's constant 
and $H$ the Hubble parameter during inflation, the corrections are small -- but, at least in 
principle, detectable \cite{Huffenberger:2004gm,Komatsu:2008hk,Bernui:2008ei}.

Since these statistical properties are a generic prediction of
(essentially) all inflationary models, they can also be inferred from two
ingredients that are usually assumed as a first approximation to our
Universe. First, since inflation was designed to stretch our Universe until
it became spatially homogeneous and isotropic, it is reasonable to expect
that all statistical momenta of the CMB should be spatially homogeneous
and rotationally invariant, regardless of their general form. 
Second, in linear perturbation theory \cite{Mukhanov:1990me} where we have
a large number of cosmological fluctuations evolving independently, we can
expect, based on the \textit{central limit theorem}, that the Universe will
obey a Gaussian distribution.

The power of this program lies, therefore, in its simplicity: if the
Universe is indeed Gaussian, homogeneous and statistically isotropic (SI), 
then essentially all the information about inflation and the linear 
(low redshift) evolution of the Universe is encoded in the variance, 
or two-point correlation function, of large-scale cosmological 
structures and/or the CMB. As it turns out, the five year dataset from the
Wilkinson Microwave anisotropy probe (WMAP) strongly supports these
predictions \cite{Hinshaw:2008kr,Komatsu:2008hk}. Moreover, the
measurements of the CMB temperature power spectrum by the WMAP team,
alongside measurements of the matter power spectrum from existing survey of
galaxies \cite{Tegmark:2003uf,Cole:2005sx} and data from type 
Ia supernovae \cite{Perlmutter:1998np,Riess:1998cb,Wood-Vasey:2007jb}, have
shown remarkable consistency with a  {\it concordance model}
($\Lambda$CDM), in which the cosmos figures as a Gaussian, spatially flat,
approximately homogeneous and statistically isotropic web of structures
composed mainly of baryons, dark matter and dark energy.

However, while the detection of a nearly scale-invariant and Gaussian
spectrum is a powerful boost to the idea of inflation, just knowing the
variance of the primordial fluctuations is not sufficient to single out
which particular inflationary model was realized in our Universe. For that
we will need not only the 2-point function, but the higher momenta of the
distribution as well. Therefore, in order to break this model degeneracy
we must go beyond the framework of the $\Lambda$CDM, Gaussian, spatially
homogeneous and statistically isotropic Universe.\\

Reconstructing our cosmic history, however, is not the only reason 
to explore further the statistical properties of the CMB. 
The full-sky temperature maps by WMAP \cite{Bennett:2003bz,Komatsu:2008hk}
have revealed the existence of a series of large-angle anomalies -- which,
incidentally were (on hindsight) already visible in the  lower-resolution
COBE data \cite{Smoot:1992td}. These anomalies suggest that at least one of
our cherished hypothesis underlying the standard cosmological model might 
be wrong -- even as a first-order approximation. Perhaps the most
intriguing anomalies (described in more detail in other review papers in
this volume) are the low value of the quadrupole and its 
alignment of the quadrupole ($\ell=2$) with the octupole ($\ell=3$) 
\cite{deOliveiraCosta:2003pu,Copi:2003kt,Schwarz:2004gk,Land:2005ad,
Copi:2005ff,Tegmark:2003ve}, the sphericity \cite{Copi:2005ff} (or lack of
planarity \cite{Abramo:2009fe}), of the multipole $\ell=5$, and the
north-south asymmetry \cite{Bernui:2005pz,Bernui:2008cr,Eriksen:2003db,Eriksen:2007pc,Pietrobon:2009qg}. 
In the framework of the standard cosmological model, these are very unlikely statistical
events, and yet the evidence that they exist in the real data (and are not
artifacts of poorly subtracted extended foregrounds -- e.g.,
\cite{Abramo:2006hs}) is strong.

Concerning theoretical explanations, even though we have by now an
arsenal of {\it ad-hoc} models designed to account for the existence of 
these anomalies, none has yet quite succeeded in explaining their 
origin. Nevertheless, they all share the point of view that the detected 
anomalies might be related to a deviation of gaussianity and/or statistical isotropy.

In this review we will describe, first, how to characterize, from
the point of view of the underlying spacetime symmetries,
both non-gaussianity and statistical anisotropy.
We will adopt two guiding principles. The first is that
gaussianity and SI, being completely different properties of a random
variable, should be treated separately, whenever possible or practical.
Second, since there is only one type of gaussianity and SI but virtually
infinite ways away from them, it is important to try to measure
these deviations without a particular model or anomaly in mind
-- although we may eventually appeal to particular models as 
illustrations or as a means of comparison.
This approach is not new and, although not usually
mentioned explicitly, it has been adopted in a number
of recent papers \cite{Abramo:2006gw,Hanson:2009gu}.

One of the main motivations for this model-independent approach is the 
difficult issue of {\it aprioristic} statistics: one can only test the
random nature of a process if it can be repeated a very large (formally,
infinite) number of times. Since the CMB only changes on a timescale of
tens of millions of years, waiting for our surface of last scattering to
probe a different region of the Universe is not a practical proposition.
Instead, we are stuck with one dataset (a sequence of apparently random
numbers), which we can subject to any number of tests. Clearly, by sheer
chance about 30\% of the tests will give a positive detection with 70\%
confidence level (C.L.), 10\% will give a positive detection with 90\%
C.L., and so on. With enough time, anyone can come up with detections of
arbitrarily high significance -- and ingenuity will surely accelerate this
process. Hence, it would be useful to have a few guiding principles to
inform and motivate our statistical tests, so that we don't
end up shooting blindly at a finite number of fish in a small
wheelbarrow. \\

This review is divided in two parts. We start Part I by reviewing
the basic statistical framework behind linear perturbation theory
(\S\ref{sec:general-structure}). This serves as a motivation for
\S\ref{sec:beyond-sm}, where we discuss the formal aspects of
non-Gaussian and statistically isotropic models
(\S\ref{sub:Non-Gaussian-and-SI}), as well as Gaussian models of
statistical anisotropy (\S\ref{sub:Gaussian-and-SA-models}). Part
II is devoted to a discussion on model-independent cosmological tests of
non-gaussianity and statistical anisotropy and their application to CMB
data. We focus on two particular tests, namely, the multipole vectors
statistics (\S\ref{sec:mv}) and functional modifications of the two-point
correlation function (\S\ref{sec:tcf}). After discussing how such tests 
are usually carried out when searching for anomalies in CMB data
(\S\ref{sec:standard-calc}), we present a new formalism which generalizes
the standard procedure by including the ergodicity of cosmological data
as a possible source of errors (\S\ref{sec:conv-prob}). This formalism is illustrated 
in \S\ref{sec:chi2-test}, where we carry a search of planar-type deviations of isotropy 
in CMB data. We then conclude in \S\ref{sec:conclusions}.

\newpage
\part{The linearized Universe}

\section{General structure}{\label{sec:general-structure}}

We start by defining the temperature fluctuation field. Since the
background radiation is known to have an average temperature of 
2.725K, we are interested only in deviations from this value at
a given direction $\hat{n}$ in the CMB sky. So let us consider the
dimensionless function on $S^2$: 
\begin{equation}
\Theta (\hat{n})\equiv\frac{T(\hat{n})-T_{0}}{T_{0}} \; ,
\label{DeltaT-def}
\end{equation}
where $T_0=2.725$ K is the blackbody temperature of the mean 
photon energy distribution -- which, if homogeneity holds, is also equal to the ensemble average
of the temperature.

In full generality, the fluctuation field is not only a function
of the position vector $\vec{n}$, but also of the time in
which our measurements are taken. In practice, the time and displacement
of measurements vary so slowly that we can ignore these dependences
altogether. Therefore, we can equally well consider this function as one 
defined only on the unit radius sphere $S^2$, for which the following decomposition 
holds: 
\begin{equation}
\Theta(\hat{n})=\sum_{\ell,m}a_{\ell m}Y_{\ell m}(\hat{n})\,.
\label{DeltaT}
\end{equation}
Since the spherical harmonics $Y_{\ell m} (\hat{n})$ obey
the symmetry $Y_{\ell m}^* (\hat{n}) = (-1)^{m} Y_{\ell,-m} (\hat{n})$,
the fact that the temperature field is a real function implies
the identity $a_{\ell m}^{*}=(-1)^{m}a_{\ell,-m}$. This
means that each temperature multipole $\ell$ is completely
characterized by $2\ell+1$ real degrees of freedom.

\subsection{From inhomogeneities to anisotropies: linear theory
\label{sub:theory}}

The ultimate source of anisotropies in the Universe are the
inhomogeneities in the baryon-photon fluid, as well as their 
associated spacetime metric fluctuations. If the photons
were in perfect equilibrium with the baryons up to a sharply defined
moment in time (the so-called instant recombination approximation),
their distribution would have only one parameter (the equilibrium
temperature at each point), so that photons flying off in any direction
would have exactly the same energies.
In that case, the photons we see today coming from a line-of-sight $\hat{n}$
would reflect simply the density and gravitational potentials (the ``sources'') at 
the position $R \, \hat{n}$, where $R$ is the radius to that 
(instantaneous) last scattering surface.
Evidently, multiple scatterings at the epoch of recombination,
combined with the fact that anisotropies themselves act as
sources for more anisotropies, complicate this picture,
and in general the relationship of the sources with the
anisotropies must be calculated from either a set of Einstein-Boltzmann
equations or, equivalently, from the line-of-sight
integral equations coupled with the Einstein, continuity
and Euler equations \cite{Dodelson:03}.

Assuming for simplicity that recombination was instantaneous, at a time $\eta_R$,
the linear kernels of Eq. (\ref{sources}) reduce to 
$K_i(\vec{x}\, ',\eta '; \hat{n}) \rightarrow \beta_i \delta(\eta'-\eta_R) 
\delta(\vec{x}\, '- \hat{n} R)$, where $R=\eta_0-\eta_R$ and $\beta_i$
are constant coefficients. The photon distribution that we measure on Earth 
would therefore be given by:
\begin{equation}
\label{sources_2}
\Theta(\hat{n}) \approx \sum_i \beta_i S^i(\vec{x} \, ' = \hat{n}R,\eta'=\eta_R)  \; .
\end{equation}
We can also express this result in terms of the Fourier spectrum of the sources:
\begin{equation}
\label{sources_3}
\Theta(\hat{n}) \approx 
\sum_i \beta_i \int \frac{d^3 k}{(2\pi)^3} e^{i \vec{k} \cdot \hat{n} R} S^i(\vec{k},\eta_R)  \; .
\end{equation}
Now we can use what is
usually referred to as ``Rayleigh's expansion'' (though Watson, in
his classic book on Bessel functions, attributes this to Bauer, J. f. Math. LVI, 1859):
\begin{equation}
\label{Raleigh}
e^{i \vec{k} \cdot \vec{x}}
=  4\pi \sum_{\ell m} i^\ell \, j_\ell (k x) \, Y^*_{\ell m} (\hat{k}) \,
Y_{\ell m} (\hat{x}) \; ,
\end{equation}
where $j_\ell(z)$ are the spherical Bessel functions.
Substituting Eq. (\ref{Raleigh}) into Eq. (\ref{sources_3}) we
obtain that:
\begin{equation}
\label{Falm}
a_{\ell m} = \int d^2 \hat{n} \, Y^*_{\ell m} (\hat{n}) \, \Theta(\hat{n}) 
\approx \int \frac{d^3 k}{(2\pi)^3}
\, \Theta(\vec{k})\,\times4\pi\,i^\ell\,j_\ell(kR)\,Y_{\ell m}^*(\hat{k}) \; ,
\end{equation}
where we have loosely collected the sources into the term 
$\Theta(\vec{k}) \equiv \sum_i \beta_i S^i (\vec{k},\eta_R)$.
This expression conveys well the simple
relation between the Fourier modes and the spherical harmonic
modes. Therefore, up to coefficients which are known given
some background cosmology, the statistical
properties of the harmonic coefficients $a_{\ell m}$
are inherited from those of the Fourier modes $\Theta(\vec{k})$
of the underlying matter and metric fields.
Notice that the properties of the $a_{\ell m}$'s under rotations, on
the other hand, have nothing to do with the statistical
properties of the fluctuations: they come directly from the 
spherical harmonic functions $Y_{\ell m}$.

\subsection{Statistics in Fourier space}

The characterization of the statistics of random variables is 
most commonly expressed in terms of the correlation functions. 
The two-point correlation function is the ensemble expectation value:
\be
\label{2ptf}
C(\vec{k},\vec{k}') \equiv
\langle \Theta (\vec{k}) \Theta(\vec{k}') \rangle \; .
\ee
In the absence of any symmetries, this would be a generic function 
of the arguments $\vec{k}$ and $\vec{k}'$, with only two constraints:
first, because $\Theta(\vec{x})$ is a real function, 
$\Theta^*(\vec{k})=\Theta(-\vec{k})$, hence in our definition
$C^*(\vec{k},\vec{k}')=C(-\vec{k},-\vec{k}')$; 
and second, due to the associative nature of the expectation 
value, $C(\vec{k},\vec{k}')=C(\vec{k}',\vec{k})$. It is obvious how
to generalize this definition to 3, 4 or an arbitrary number of
fields at different $\vec{k}$'s (or ``points''.)

Let us first discuss the issue of gaussianity. If we say that the
variables $\Theta(\vec{k})$ are Gaussian random numbers,
then all the information that characterizes their distribution is 
contained in their two-point function $C(\vec{k},\vec{k}')$.
The probability distribution function (pdf) is then formally given by:
$$
P[\Theta(\vec{k}),\Theta(\vec{k}')] \sim 
\exp \left[-\frac{\Theta(\vec{k}) \Theta(\vec{k}')}{2 \, C(\vec{k},\vec{k}') } 
\right] \; .
$$
In this case, all higher-order correlation functions are either 
zero (for odd numbers of points) or they are simply 
connected to the two-point function by means of Wick's Theorem:
\be
\label{Wick}
\langle \Theta(\vec{k}_1) \Theta(\vec{k}_2) \ldots \Theta(\vec{k}_{2N})
\rangle_{\rm G}
= \sum_{i,j} \prod_{\alpha=1}^N
B_{i,j}^\alpha \langle \Theta(\vec{k}_{i}) \Theta(\vec{k}_{j}) \rangle \; ,
\ee
where the sum runs over all permutations of the pairs of wave vectors
and $B_{i,j}$ are weights.

Second, let's consider the issue of homogeneity. 
A field is homogeneous if its expectation 
values (or averages) do not dependent on the spatial points
where they are evaluated. In terms of the $N$-point functions in 
real space, we should have that:
\be
\label{Npfr}
\langle \Theta(\vec{x}_1) \Theta(\vec{x}_2) \ldots \Theta(\vec{x}_N)
\rangle \overset{\rm Homog.}{\rightarrow} C_N
(\vec{x}_1-\vec{x}_2, ..., \vec{x}_{N-1} - \vec{x}_N )  \; .
\ee
Writing this expression in terms of the Fourier modes, we get that:
\begin{eqnarray}
\nonumber
\langle \Theta(\vec{x}_1) \Theta(\vec{x}_2) \ldots \Theta(\vec{x}_N) \rangle 
&=& \int \frac{d^3 k_1 \, d^3 k_1 \, \ldots d^3 k_N \, }{(2\pi)^{3N}}
e^{-i \vec{k}_1\cdot \vec{x}_1}
e^{-i \vec{k}_2\cdot \vec{x}_2}
\ldots
e^{-i \vec{k}_N\cdot \vec{x}_N}
\\ \label{Npfs}
& \times &
\langle \Theta(\vec{k}_1) \Theta(\vec{k}_2) \ldots \Theta(\vec{k}_N) \rangle 
\; .
\end{eqnarray}
Homogeneity demands that the expression in Eq. (\ref{Npfs}) 
is a function of the {\it distances} between spatial 
points only, not of the points themselves. Hence, 
the expectation value in Fourier space on the
right-hand-side of this expression must be proportional to 
$\delta(\vec{k}_1+\vec{k}_2+ \ldots + \vec{k}_N)$. In other words,
the hypothesis of homogeneity constrains the $N$-point function
in Fourier space to be of the form:
\be
\label{Npfk}
\langle \Theta(\vec{k}_1) \Theta(\vec{k}_2) \ldots \Theta(\vec{k}_N) 
\rangle_{\rm H}
= (2\pi)^3 \tilde{N} (\vec{k}_1,\vec{k}_2,\ldots,\vec{k}_N) \, 
\delta(\vec{k}_1+\vec{k}_2+ \ldots + \vec{k}_N) \; .
\ee
Notice that the ``$(N-1)$-spectrum'' in Fourier space, $\tilde{N}$, 
can still be a function of the directions of the wavenumbers 
$\vec{k}_i$ (it will be, in fact, a function of $N-1$ such
vectors, due to the global momentum conservation expressed
by the $\delta$-function.) Models which realize the
general idea of Eq. (\ref{Npfk}) correspond to homogeneous but
anisotropic universes 
\cite{Pullen:2007tu,Pereira:2007yy,Pitrou:2008gk,Gumrukcuoglu:2007bx}.

There is a useful diagrammatic illustration for the
$N$-point functions in Fourier space that enforce homogeneity.
Notice that we could use the $\delta$-function in Eq. (\ref{Npfk}) 
to integrate out any one of the momenta $\vec{k}_i$ in Eq. (\ref{Npfs}).
Let us instead rewrite the $\delta$-functions in terms of triangles, so
for the 4-point function we have:
\be
\label{deltas}
\delta(\vec{k}_1+\vec{k}_2+\vec{k}_3+\vec{k}_4)
= \int \, d^3 q \, \delta(\vec{k}_1+\vec{k}_2-\vec{q})
\,  \delta(\vec{k}_3+\vec{k}_4+\vec{q}) \; ,
\ee
whereas for the 5-point function we have:
\be
\label{deltas2}
\delta(\vec{k}_1+\vec{k}_2+\vec{k}_3+\vec{k}_4+\vec{k}_5)
= \int \, d^3 q \, d^3 q' \, \delta(\vec{k}_1+\vec{k}_2-\vec{q})
\, \delta(\vec{q}+\vec{k}_3-\vec{q} \, ') 
\,  \delta(\vec{q} \, '+\vec{k}_4+\vec{k}_5) \; ,
\ee
and so on, so that the $N$-point $\delta$-function is
reduced to $N-2$ triangles with $N-3$ ``internal momenta''
(the idea is nicely illustrated in Fig. 1.)
Substituting the expression for the $N$-point $\delta$-function 
into Eq. (\ref{Npfs}) and integrating out all external momenta 
but the first ($\vec{k}_1$) and last ($\vec{k}_N$), the result is that:
\begin{eqnarray}
\nonumber
\langle \Theta(\vec{x}_1) \Theta(\vec{x}_2) \ldots \Theta(\vec{x}_N) \rangle
&=& \frac{1}{(2\pi)^{3N}} 
\int  d^3 k_1 \, d^3 q_1 \ldots d^3 q_{N-3} \, d^3 k_N
\\ \nonumber
&\times& e^{i \vec{k}_1 \cdot (\vec{x}_1-\vec{x}_2)}
e^{ i \vec{q}_1 \cdot (\vec{x}_2-\vec{x}_3)}
\ldots
e^{i \vec{q}_{N-3} \cdot (\vec{x}_{N-2}-\vec{x}_{N-1})}
e^{i \vec{k}_N \cdot (\vec{x}_{N-1}-\vec{x}_N)} 
\\ \label{Ndiag}
& \times & 
\langle 
\Theta(\vec{k}_1) 
\Theta(\vec{q}_1-\vec{k}_1) 
\ldots 
\Theta(\vec{k}_N) 
\rangle \; .
\end{eqnarray}
This expressions shows explicitly that the real-space $N$-point
function above does not depend on any particular spatial point,
only on the intervals between points.

\begin{center}
\begin{figure}[h]
\begin{centering}
\includegraphics[scale=0.5]{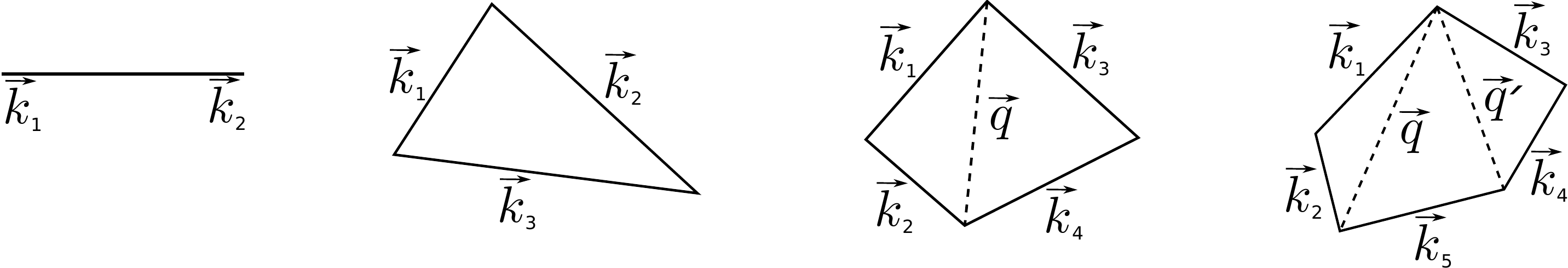}
\par\end{centering}
\caption{Diagrammatic representation of the 2, 3, 4 and 5-point correlation
functions in Fourier space. The dashed lines represent internal momenta.}
\label{npoint-diagrams-k}
\end{figure}
\par\end{center}

Finally, what are the constraints imposed on the $N$-point functions that
come from isotropy alone? Clearly, no dependence on the {\it directions}
defined by the points, $\vec{x}_i-\vec{x}_j$, can arise in the final 
expression for the $N$-point functions in real space, so from Eq. 
(\ref{Npfs}) we see that the $N$-point function in Fourier space
should depend only on the moduli of the wavenumbers -- up to
some momentum-conservation $\delta$-functions, which
naturally carry vector degrees of freedom.

In this review we will mostly be concerned with tests of isotropy
given homogeneity (but not necessarily Gaussianity), so in our case
we will usually assume that the $N$-point function in Fourier space 
assumes the form given in Eq. (\ref{Npfk}).

\subsection{Statistics in harmonic space}\label{sec:statistics-harm-space}

In the previous section we characterized the statistics of our field
in Fourier space, which in most cases is most easily related to
fundamental models such as inflation. Now we will change to
harmonic representation, because that's what is most directly
related to the observations of the CMB, $\Theta(\hat{n})$, 
which are taken over the unit sphere $S^2$.
We will discuss mostly the two-point function here, and we
defer a fuller discussion of $N$-point functions in harmonic
space to Section 3.

From Eq. (\ref{Falm}) we can start by taking the two-point
function in harmonic space, and computing it in terms of the
two-point function in Fourier space:
\be
\label{2pth}
\langle a_{\ell m} a^*_{\ell' m'} \rangle 
= \int \frac{d^3 k \, d^3 k'}{(2\pi)^6} \, (4\pi)^2 \, i^{\ell} (-i)^{\ell'} \,
j_\ell(kR) \, j_{\ell'} (k'R) \, Y_{\ell m} (\hat{k}) Y^*_{\ell' m'} (\hat{k}') \,
\langle \Theta(\vec{k}) \Theta^*(\vec{k}') \rangle \; .
\ee
Under the hypothesis of homogeneity, this expression
simplifies considerably, leading to:
\be
\label{2pthh}
\langle a_{\ell m} a^*_{\ell' m'} \rangle_{\rm H}
= \int \, d^3 k \, \frac{2}{\pi} \, i^{\ell} (-i)^{\ell'} \,
j_\ell(kR) \, j_{\ell'} (kR) \, Y_{\ell m}(\hat{k}) Y_{\ell' m'}^*(\hat{k}) \,
\times \, \tilde{N}_2(\vec{k}) \; .
\ee
If, in addition to homogeneity, we also assume isotropy, then
$\tilde{N}_2 \rightarrow P(k)$, and
the integration over angles factors out, leading to the orthogonality
condition for spherical harmonics:
$$
\int \, d^2 \hat{k} \, Y_{\ell m} (\hat{k})  \, Y^*_{\ell' m'} (\hat{k}) =
\delta_{\ell \ell'} \, \delta_{m m'} \; ,
$$
and as a result the covariance of the $a_{\ell m}$'s becomes diagonal:
\begin{eqnarray}
\label{2pCl}
\langle a_{\ell m} a^*_{\ell' m'} \rangle_{\rm H,I} 
&=& \delta_{\ell \ell'} \, \delta_{m m'}
\int \, \frac{dk}{k} \, j_\ell^2(kR) \, \frac{2}{\pi} k^3 P(k) 
\\ \nonumber
&=& 4\pi \, \delta_{\ell \ell'} \, \delta_{m m'}
\int \, d \, \log k \, j_\ell^2(kR) \, \Delta_T^2(k) 
\\ \nonumber
&\equiv& C_\ell \, \delta_{\ell \ell'} \, \delta_{m m'} \; ,
\end{eqnarray}
where we have defined the usual {\it temperature power spectrum} 
$\Delta_T(k) = k^3 P(k)/2\pi^2$ in the middle line, and the
angular power spectrum $C_\ell$ in the last line of Eq. (\ref{2pCl}). 
As a pedagogical note, let's recall that
the power spectrum basically expresses how much power 
the two-point correlation function has per unit $\log k$: 
\be
\label{TPS}
\langle \Theta(\vec{x}) \Theta(\vec{x}') \rangle_{\rm H,I}
=
\int d\log k \, \frac{\sin(k|\vec{x}-\vec{x}'|)}{k|\vec{x}-\vec{x}'|} 
\, \Delta_T^2 (k) \; .
\ee

In an analogous manner to what was done above, we can also construct the 
{\it angular} two-point correlation function in harmonic space:
\begin{eqnarray}
\label{APS}
\langle \Theta(\hat{n})  \Theta(\hat{n}') \rangle
&=& \sum_{\ell m} \sum_{\ell' m'} 
\langle a_{\ell m} a^*_{\ell' m'} \rangle 
Y_{\ell m}(\hat{n})
Y^*_{\ell m}(\hat{n}') \; .
\end{eqnarray}
The hypothesis of homogeneity by itself does not lead to
significant simplifications, but isotropy
leads to a very intuitive expression for the angular two-point function:
\begin{eqnarray}
\label{APShi}
\langle \Theta(\hat{n})  \Theta(\hat{n}') \rangle_{\rm H,I}
&=& \sum_{\ell m} \sum_{\ell' m'} 
C_\ell \, \delta_{\ell \ell'} \delta_{m m'} 
Y_{\ell m}(\hat{n})
Y^*_{\ell m}(\hat{n}') 
\\ \nonumber
&=& \sum_\ell C_\ell \frac{2\ell+1}{4\pi} P_\ell (\hat{n}\cdot \hat{n}')
\; .
\label{C-theta}
\end{eqnarray}
Clearly, not only is this expression the analogous in $S^2$ of 
Eq. (\ref{TPS}), but in fact the Fourier power spectrum $\Delta_T^2 (k)$ 
and the angular power spectrum $C_\ell$ are defined in terms of each 
other as indicated in Eq. (\ref{APShi}):
\be
\label{Cl_DT}
C_\ell = 4\pi \int d\log k \, j_\ell^2 (kR) \, \Delta_T^2 (k) \; .
\ee
Now, using the facts that the spherical Bessel function of
order $\ell$ peaks when its argument is approximately given
by $\ell$, and that $\int d\log z \, j_\ell^2(z) = 1/(2\ell(\ell+1))$,
we obtain that\footnote{This is one type of what has become 
known in the astrophysics literature as {\it Limber's approximations.}}:
\be
\label{Cl}
C_\ell \approx \frac{2\pi}{\ell(\ell+1)} \Delta_T^2(k=\ell/R) \; .
\ee
Incidentally, from this expression it is clear why it is customary to
define: 
$$
{\cal{C}}_\ell \equiv \frac{\ell(\ell+1)}{2\pi} C_\ell
\approx \Delta_T^2(k=\ell/R) \; .
$$

Using Eq. (\ref{Npfs}) we can easily generalize the results of this
subsection to $N$-point functions in $S^2$ and in harmonic space, however,
the assumption of isotropy alone does very little to simplify our life.
The hypothesis of homogeneity, on the other hand, greatly simplifies
the angular $N$-point functions, and most of the work in statistical
anisotropy of the CMB that goes beyond the two-point function assumes that
homogeneity holds. Notice that the issue of gaussianity is, as always,
confined to the question of whether or not the two-point function holds all
information about the distribution of the relevant variables, and is
therefore completely separated from questions about homogeneity and/or
isotropy.

Also notice that the separable nature of the definition (\ref{APS})
implies here as well, like in Fourier space, a reciprocity relation 
for the correlation function: 
\begin{equation}
C(\hat{n}_{1},\hat{n}_{2})=C(\hat{n}_{2},\hat{n}_{1})\,.\label{rr}
\end{equation}
This symmetry must hold regardless of underlying models, and is important
in order to analyze the symmetries of the correlation function, as we shall
see later.

Before we move on, it is perhaps important to mention that the
decomposition (\ref{APS}) is not unique. In fact, instead of the angular
momenta of the parts, $(\ell_1, m_1; \ell_2,m_2)$, we could equally well
have used the basis of total angular momentum $(L,M;\ell_1,\ell_2)$ and
decomposed that expression as:
\be
\label{C-bipolar}
C(\hat{n}_{1},\hat{n}_{2})=\sum_{L,M}\sum_{\ell_{1},\ell_{2}}
\mathcal{A}_{\ell_{1}\ell_{2}}^{LM}\mathcal{Y}_{\ell_{1}\ell_{2}}^{LM}
(\hat{n}_{1},\hat{n}_{2})
\ee
where $\mathcal{Y}_{\ell_{1}\ell_{2}}^{LM}$ are known as the 
bipolar spherical harmonics, defined by \cite{Varshalovich:1988ye}:
\[
\mathcal{Y}_{\ell_{1}\ell_{2}}^{LM}(\hat{n}_{1},\hat{n}_{2})=
[Y_{\ell_{1}}(\hat{n}_{1})\otimes Y_{\ell_{2}}(\hat{n}_{2})]_{LM} \; ,
\]
where $L $ and $M=m_1+m_2$ are the eigenvalues of the total 
and azimuthal angular momentum operators, respectively. 
This decomposition is completely equivalent to Eq. (\ref{APS}), and we can 
exchange from one decomposition to another by using the relation:
\begin{equation}
\mathcal{A}_{\ell_{1}\ell_{2}}^{LM}=\sum_{m_{1}m_{2}}
\langle a_{\ell_{1}m_{1}}a_{\ell_{2}m_{2}}\rangle
(-1)^{M+\ell_1-\ell_2}\sqrt{2L+1}
\left(\begin{array}{ccc}\ell_{1} & \ell_{2} & L\\
m_{1} & m_{2} & -M\end{array}\right)\,,
\label{bips}
\end{equation}
where the $3\times2$ matrices above are the well-known 3-j coefficients. At
this point, it is only a matter of mathematical convenience whether we choose to
decompose the correlation function as in (\ref{APS}) or as in (\ref{C-bipolar}).
Although the bipolar harmonics behave similarly to the usual spherical harmonics
in many aspects, the modulations of the correlation function as described in
this basis have a peculiar interpretation. We will not go further into detail
about this decomposition  here, as it is discussed at length in another
review article in this volume.

\subsection{Estimators and cosmic variance}\label{sec:estimators-and-cv}

Returning to the covariance matrix (\ref{2pCl}), we see that,
if we assume gaussianity of the $a_{\ell m}$'s, then the angular
power spectrum suffices to describe statistically how much the temperature
fluctuates in any given angular scale; all we have to do is to calculate
the average (\ref{2pCl}). This can be a problem, though,
since we have only one Universe to measure, and therefore only one 
set of $a_{\ell m}$'s. In other words, the average in (\ref{2pCl})
is poorly determined. 

At this point, the hypothesis that our Universe is spatially homogeneous
and isotropic at cosmological scales comes not only as simplifying
assumption about the spacetime symmetries, but also as a remedy to
this unavoidable smallness of the working cosmologist's sample space. 
If isotropy holds, different cosmological scales
are statistically independent, which means that 
we can take advantage of the ergodic
hypothesis and trade averaging over an ensemble for averaging over
space. In other words, for a given $\ell$ we can consider each of the $2\ell+1$ real
numbers in $a_{\ell m}$ as statistically independent Gaussian random
variables, and define a \textit{statistical estimator} for their variances
as the average: 
\begin{equation}
\widehat{C}_{\ell}\equiv\frac{1}{2\ell+1}\sum_{m=-\ell}^{\ell}|a_{\ell m}|
^{2} \; .
\label{Cl-hat}
\end{equation}
The smaller the angular scales ($\ell$ bigger), the larger the number
of independent patches that the CMB sky can be divided into. Therefore,
in this limit we should have:
\[
\lim_{\ell\rightarrow\infty}\widehat{C}_{\ell}=C_{\ell} \; .
\]
On the other hand, for large angular scales (small $\ell$'s), the
number of independent patches of our Universe becomes smaller, and
(\ref{Cl-hat}) becomes a weak estimation of the $C_{\ell}$'s. This
means that any statistical analysis of the Universe on large scales will
be plagued by this intrinsic \textit{cosmic sample variance}. Notice
that this is an unavoidable limit as long as we have only one observable
Universe.

Finally, it is important to keep in mind the clear distinction between the
angular power spectrum $C_{\ell}$ and its estimator (\ref{Cl-hat}).
The former is a theoretical variable which can be calculated from
first principles, as we have shown in $\S$\ref{sub:theory}. The
latter, being a function of the data, is itself a random variable.
In fact, if the $a_{\ell m}$'s are Gaussian, then we can rewrite
expression (\ref{Cl-hat}) as:
\[
\frac{(2\ell+1)}{C_{\ell}}\widehat{C}_{\ell}=X_\ell\,,\qquad X_\ell=\sum_{m=-\ell}^{\ell}
\frac{|a_{\ell m}|^{2}}{C_{\ell}} \; ,
\]
where $X_\ell$ is a chi-square random variable with $2\ell+1$ degrees
of freedom. According to the central limit theorem, when 
$\ell \rightarrow \infty$, $X_\ell$ approaches a standard normal
variable\footnote{A standard normal variable is a Gaussian variable $X$ 
with zero mean and unit variance. Any other Gaussian variable $Y$ with mean
$\mu$ and variance $\sigma$ can be obtained from $X$ through $Y=\sigma
X+\mu$.}, which implies that $\widehat{C}_{\ell}$ will itself follow a
Gaussian distribution. Its mean can be easily calculated using (\ref{2pCl})
and (\ref{Cl-hat}), and is of course given by:
\[
\langle\widehat{C}_{\ell}\rangle=C_{\ell} \; ,
\]
which shows that the $\widehat{C}_\ell$'s are {\it unbiased} estimators of 
the $C_{\ell}$'s.
It is also straightforward to calculate its variance (valid for any $\ell$):
\[
\langle(\widehat{C}_{\ell}-C_{\ell})(\widehat{C}_{\ell'}-C_{\ell'})\rangle=
\frac{2}{2\ell+1}\, C_{\ell}^{2}\;\delta_{\ell\ell'} \; .
\]
Because this estimator does not couple different cosmological scales,
it has the minimum \textit{cosmic variance} we can expect from an
estimator due to the finiteness of our sample -- so it is {\it optimal} in
that sense. $\widehat{C}_{\ell}$
is therefore the best estimator we can build to measure the statistical
properties of the multipolar coefficients $a_{\ell m}$ when both
statistical isotropy and gaussianity hold.

In later Sections we will explore angular or harmonic $N$-point functions
for which the assumption of isotropy does not hold. However, it is
important to remember at all times that we have only one map, which means
one set of $a_{\ell m}$'s. The estimator for the angular power spectrum, 
$\widehat{C}_\ell$, takes into account {\it all} the $a_{\ell m}$'s by
dividing them into the different $\ell$'s and summing over all
$m\in(-\ell,\ell)$. Clearly, it will inherit a sample variance for small
$\ell$'s, when the $a_{\ell m}$'s can only be divided into a few
``independent parts''. As we try to estimate higher-order objects such as
the $N$-point functions, we will have to subdivide the $a_{\ell m}$'s into
smaller and smaller subsamples, which are not necessarily independent (in
the statistical sense) of each other. So, the price to pay for aiming at
higher-order statistics is a worsening of the cosmic sample variance.

\subsection{Correlation and statistical independence
\label{correlation}}

The covariance given in Eq. (\ref{2pCl}) has two distinct,
important properties. First, note that its diagonal entries, the 
$C_{\ell}$'s, are $m$-independent coefficients; this is crucial for having
statistical isotropy, as we will show latter. Second, statistical isotropy
at the Gaussian level implies that different cosmological ``scales'' (understood 
here as meaning the modes with total angular momentum $\ell$ and azimuthal momentum
$m$) should be \textit{statistically independent} of each other -- and 
this is represented by the Kronecker deltas in (\ref{2pCl}). 

In fact, statistical independence of cosmological scales is a particular
property of Gaussian and statistically isotropic random fields, and is not
guaranteed to hold when gaussianity is relaxed. We will see in the next
Section that the rotationally invariant 3-point correlation function (and
in general any $N>2$ correlation function) couples to the three scales
involved. In particular, if it happens that the Gaussian contribution of
the temperature field is given by (\ref{2pCl}), but at least one of its
non-Gaussian moments are nonzero, then the fact that a particular
correlation is zero, like for example 
$\langle a_{2m_{1}}a_{3m_{2}}^{*}\rangle$,
does not imply that the scales $\ell=2$ and $\ell=3$ are (statistically)
independent. This is just a restatement of the fact that, while statistical
independence implies null correlation, the opposite is not necessarily
true. This can be illustrated by the following example: consider a random
variable $\alpha$ distributed as:
\[
P(\alpha)=\begin{cases}
1 & \alpha\in[0,1]\\
0 & \mbox{otherwise}\,.\end{cases}
\]
Let us now define two other variables $x=\cos(2\pi\alpha)$ and $y=\sin(2\pi\alpha)$.
From these definitions, it follows that $x$ and $y$ are statistically
dependent variables, since knowledge of the mean/variance of $x$
automatically gives the mean/variance of $y$. However, these variables
are clearly uncorrelated:
\[
\langle xy\rangle=\frac{1}{2\pi}\int_{0}^{2\pi}\cos\eta\sin\eta d\eta=0\,.
\]
Although correlations are among cosmologist's most popular tools when
analyzing CMB properties, statistical independence may turn out to
be an important property as well, specially at large angular scales,
where cosmic variance is more of a critical issue.\\

\section{Beyond the standard statistical model}{\label{sec:beyond-sm}}

Until now we have been analyzing the properties of Gaussian and 
statistically isotropic random temperature fluctuations. This gives us a
fairly good statistical description of the Universe in its linear regime,
as confirmed by the astonishing success of the $\Lambda$CDM model.
This picture is incomplete though, and we have good reasons to search
for deviations of either gaussianity and/or statistical isotropy.
For example, the observed clustering of matter in galactic environments
certainly goes beyond the linear regime where the central limit theorem
can be applied, therefore leading to large deviations of gaussianity
in the matter power spectrum statistics. Besides, deviations
of the cosmological principle may leave an imprint in
the statistical moments of cosmological observables, which can be
tested by searching for spatial inhomogeneities \cite{Uzan:2008qp} 
or directionalities \cite{Prunet:2004zy}. 

But how do we plan to go beyond the standard model, given that there
is only one Gaussian and statistically isotropic description of the
Universe, but infinite possibilities otherwise? This is in fact an
ambitious endeavor, which may strongly depend on observational and
theoretical hints on the type of signatures we are looking for. In
the absence of extra input, it is important to classify these signatures
in a general scheme, differentiating those which are non-Gaussian
from those which are anisotropic. Furthermore, given that the signatures
of non-gaussianity may in principle be quite different from that of statistical
anisotropy, such a classification is crucial for data analysis, which requires 
sophisticated tools capable of separating these two issues\footnote{Although 
gaussianity and homogeneity/isotropy are mathematically 
distinct properties, it is possible for a Gaussian but inhomogeneous/anisotropic 
model to look like an isotropic and homogeneous non-gaussian model. See for 
example \cite{Ferreira:1997wd}.}.

We therefore start $\S$\ref{sub:Non-Gaussian-and-SI} by analyzing deviations of 
gaussianity when statistical isotropy holds. In $\S$\ref{sub:Gaussian-and-SA-models} 
we keep the hypothesis of gaussianity and analyze the consequences of breaking 
statistical rotational invariance.

\subsection{Non-Gaussian and SI models\label{sub:Non-Gaussian-and-SI}}

\subsubsection{Rotational invariance of $N$-point correlation functions.}
\label{sec:Invariant-N-point}

We turn know to the question of non-Gaussian but statistically isotropic
probabilities distributions. We will keep working with the $N$-point
correlation function defined in harmonic space, 
\begin{equation}
\langle a_{\ell_{1}m_{1}}a_{\ell_{2}m_{2}}\dots a_{\ell_{N}m_{N}}\rangle
\label{harmonic-npoint}
\end{equation}
since knowledge of these functions enables one to fully reconstruct the CMB
temperature probability distribution. Specifically, we would like to know
the form of any $N$-point correlation function which is invariant under
arbitrary 3-dimensional spatial rotations. When rotated to a new (primed)
coordinate system, the $N$-point correlation function transforms as:
\begin{equation}
\langle a_{\ell_{1}m'_{1}}a_{\ell_{2}m'_{2}}\dots a_{\ell_{N}m'_{N}}
\rangle=\sum_{{\rm all}\, m}\langle a_{\ell_{1}m_{1}}a_{\ell_{2}m_{2}}\dots
a_{\ell_{N}m_{N}}\rangle
D_{m'_{1}m_{1}}^{\ell_{1}}D_{m'_{2}m_{2}}^{\ell_{2}}\dots
D_{m'_{N}m_{N}}^{\ell_{n}} \, ,
\label{N-point-SI}
\end{equation}
where the $D_{m_{i}m_{i}'}^{\ell}(\alpha,\beta,\gamma)$'s are the
coefficients of the Wigner rotation-matrix, which depend on the three
Euler-angles $\alpha$, $\beta$ and $\gamma$ characterizing the
rotation. Notice that in this notation the primed (rotated) system is
indicated by the primed $m$'s.
For the 2-point correlation function, we have already seen
that the well-known expression 
\[
\langle a_{\ell_{1}m_{1}}a_{\ell_{2}m_{2}}\rangle=(-1)^{m_{2}}C_{\ell_{1}}
\delta_{\ell_{1}\ell_{2}}\delta_{m_{1},-m_{2}}
\]
does the job: 
\begin{eqnarray*}
\langle a_{\ell_{1}m'_{1}}a_{\ell_{2}m'_{2}}\rangle & = &
C_{\ell_{1}}\left(\sum_{m_{1}}(-1)^{m_{1}}D_{m'_{1}m_{1}}^{\ell_{1}}
D_{m'_{2}-m_{1}}^{\ell_{1}}\right)\delta_{\ell_{1}\ell_{2}}\\
 & = &
(-1)^{m'_{2}}C_{\ell_{1}}\delta_{\ell_{1}\ell_{2}}\delta_{m'_{1},-m'_{2}}\,.
\end{eqnarray*}
Note the importance of the \textit{angular spectrum}, $C_{\ell}$, being a
$m$-independent function. 

What about the 3-point function? In this case,
the invariant combination is found to be:
\[
\langle
a_{\ell_{1}m_{1}}a_{\ell_{2}m_{2}}a_{\ell_{3}m_{3}}\rangle=B_{\ell_{1}\ell_{2}
\ell_{3}}
\left(\begin{array}{ccc}
\ell_{1} & \ell_{2} & \ell_{3}\\
m_{1} & m_{2} & m_{3}\end{array}\right)
\]
which can be verified by straightforward calculations. Again, the
non-trivial physical content of this statistical moment is contained
in an arbitrary but otherwise $m$-independent function: the \textit{bispectrum}
$B_{\ell_{1}\ell_{2}\ell_{3}}$ \cite{Fry:1985ri,Luo:1993xx,Spergel:1999xn}. 
As we anticipated in Section 2,
rotational invariance of the 3-point correlation is not enough to
guarantee statistical independence of the three cosmological scales
involved in the bispectrum, although in principle a particular model
could be formulated to ensure that $B_{\ell_{1}\ell_{2}\ell_{3}}\propto\delta_{\ell_{1}\ell_{2}}\delta_{\ell_{2}\ell_{3}}$,
at least for some subset of a general geometric configuration of the
3-point correlation function. 

These general properties hold for all the $N$-point correlation function.
For the 4-point correlation function, for example, Hu \cite{Hu:2001fa}
have found the following rotationally invariant combination
\[
\langle a_{\ell_{1}m_{1}}\dots a_{\ell_{4}m_{4}}\rangle=\sum_{LM}Q_{\ell_{3}\ell_{4}}^{\ell_{1}\ell_{2}}(L)(-1)^{M}
\left(\begin{array}{ccc}
\ell_{1} & \ell_{2} & L\\
m_{1} & m_{2} & -M\end{array}\right)\left(\begin{array}{ccc}
\ell_{3} & \ell_{4} & L\\
m_{3} & m_{4} & M\end{array}\right)
\]
where the $Q_{\ell_{3}\ell_{4}}^{\ell_{1}\ell_{2}}(L)$ function is
known as the \textit{trispectrum}, and $L$ is an internal angular
momentum needed to ensure parity invariance. In a likewise manner,
it can be verified that the following expression
\begin{eqnarray*}
\langle a_{\ell_{1}m_{1}}\dots a_{\ell_{5}m_{5}}\rangle & = & \sum_{LM}\sum_{L'M'}P_{\ell_{3}\ell_{4}\ell_{5}}^{\ell_{1}\ell_{2}}(L,L')(-1)^{M+M'}
\left(\begin{array}{ccc}
\ell_{1} & \ell_{2} & L\\
m_{1} & m_{2} & -M\end{array}\right)\\
 &  & \qquad\qquad\times\left(\begin{array}{ccc}
\ell_{3} & \ell_{4} & L'\\
m_{3} & m_{4} & -M'\end{array}\right)\left(\begin{array}{ccc}
\ell_{5} & L & L'\\
m_{5} & M & M'\end{array}\right)
\end{eqnarray*}
gives the rotationally invariant quadrispectrum 
$P_{\ell_{3}\ell_{4}\ell_{5}}^{\ell_{1}\ell_{2}}(L,L^{'})$.

\begin{center}
\begin{figure}[h]
\begin{centering}
\includegraphics[scale=0.5]{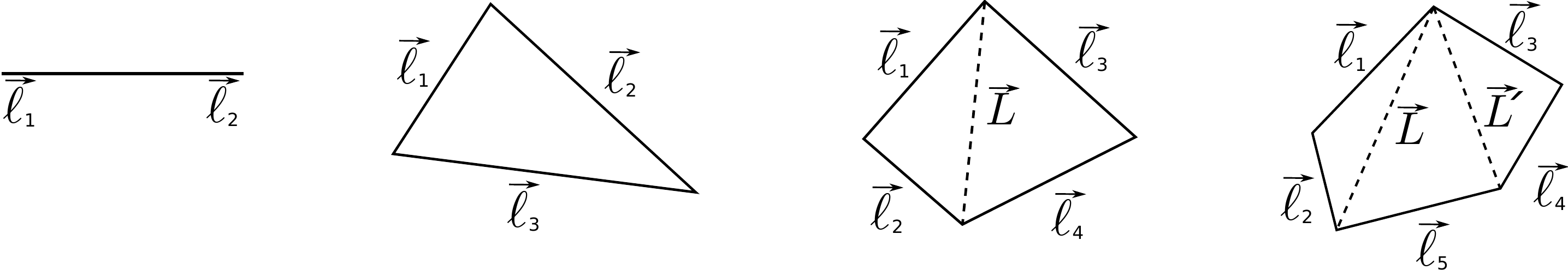}
\par\end{centering}
\caption{Diagrammatic representation of the 2, 3, 4 and 5-point correlation
functions in harmonic space. Here $\vec{\ell}$ actually represents
the pair $(\ell,m)$.}
\label{npoint-diagrams-l}
\end{figure}
\par\end{center}
The examples above should be enough to show how the general structure
of these functions emerges under SI: apart from a $m$-independent
function, every pair of momenta $\ell_{i}$ in these functions are
connected by a triangle, which in turn connects itself to other triangles
through internal momenta when more than 3 scales are present. In Fig.
\ref{npoint-diagrams-l} we show some diagrams representing the functions
above.\\

Although we have always shown $N$-point functions which are rotationally
invariant, the procedure used for obtaining them was rather intuitive,
and therefore does not offer a recipe for constructing general invariant
correlation functions. Furthermore, it does not guarantee that this
procedure can be extended for arbitrary $N$'s. Here we will present
a recipe for doing that, which also guarantees the uniqueness of the
solution.

The general recipe for obtaining the rotationally invariant $N$-point
function is as follows: from the expression (\ref{N-point-SI}) above, we
start by contracting every pairs of Wigner functions, where by
``contracting'' we mean using the identity
\begin{eqnarray*}
D_{m'_{1}m_{1}}^{\ell_{1}}(\omega)D_{m'_{2}m_{2}}^{\ell_{2}}(\omega) & = & \sum_{L,M,M'}\left(\begin{array}{ccc}
\ell_{1} & \ell_{2} & L\\
m'_{1} & m'_{2} & -M\end{array}\right)\left(\begin{array}{ccc}
\ell_{1} & \ell_{2} & L\\
m_{1} & m_{2} & -M'\end{array}\right)\\
 &  & \times(2L+1)(-1)^{M+M'}D_{M'M}^{L}(\omega)
\end{eqnarray*}
and where $\omega=\{\alpha,\beta,\gamma\}$ is a shortcut notation
for the three Euler angles. Once this contraction is done there will
remain $\lceil N/2\rceil$ $D$-functions, which can again be contracted
in pairs. This procedure should be repeated until there is only one
Wigner function left, in which case we will have an expression of
the following form:
\[
\langle a_{\ell_{1}m_{1}}\dots a_{\ell_{N}m_{N}}\rangle=\sum_{{\rm all}\, m'}
\langle a_{\ell_{1}m'_{1}}\dots a_{\ell_{N}m'_{N}}\rangle\times\sum
\mbox{geometrical factors}\times D_{MM'}^{L}(\omega)\,.
\]
Now, we see that the only way for this combination to be rotationally
invariant is when the remaining $D_{MM'}^{L}$ function above does
not depend on $\omega$, {\it i.e.},
$D_{MM'}^{L}(\omega)=\delta_{L0}\delta_{M0}\delta_{M'0}$.
Once this identity is applied to the geometrical factors, we are done,
and the remaining terms inside the primed $m$-summation will give
the rotationally invariant ($N$-1)-spectrum. 

As an illustration of this algorithm, let us construct the rotationally
invariant spectrum and bispectrum. For the 2-point function there
is only one contraction to be done, and after we simplify the last
Wigner function we arrive at
\[
\langle a_{\ell_{1}m_{1}}a_{\ell_{2}m_{2}}\rangle=\left[\sum_{m'_{1}}
\frac{\langle|a_{\ell_{1}m'_{1}}|^{2}\rangle}{2\ell_{1}+1}\right](-1)^{m_{2}}
\delta_{\ell_{1}\ell_{2}}\delta_{m_{1},-m_{2}} \; ,
\]
where, of course:
\[
C_{\ell}\equiv\frac{1}{2\ell+1}\sum_{m}\langle|a_{\ell m}|^{2}\rangle \; ,
\]
is the well-known definition of the temperature angular spectrum.
For the 3-point function there are two contractions, and the simplification
of the last Wigner function gives 
\[
\langle a_{\ell_{1}m_{1}}a_{\ell_{2}m_{2}}a_{\ell_{3}m_{3}}\rangle=
\left[\sum_{m'_{1},m'_{2},m'_{3}}\langle
a_{\ell_{1}m'_{1}}a_{\ell_{2}m'_{2}}a_{\ell_{3}m'_{3}}\rangle
\left(\begin{array}{ccc}
\ell_{1} & \ell_{2} & \ell_{3}\\
m'_{1} & m'_{2} & m'_{3}\end{array}\right)\right]\left(\begin{array}{ccc}
\ell_{1} & \ell_{2} & \ell_{3}\\
m_{1} & m_{2} & m_{3}\end{array}\right)\,.
\]
From this expression and the ortoghonality of the 3-j's symbols (see the Appendix), we can 
immediately identify the definition of the bispectrum:
\[
B_{\ell_{1}\ell_{2}\ell_{3}}\equiv\sum_{m_{1},m_{2},m_{3}}\langle
a_{\ell_{1}m_{1}}a_{\ell_{2}m_{2}}a_{\ell_{3}m_{3}}\rangle
\left(\begin{array}{ccc}\ell_{1} & \ell_{2} & \ell_{3}\\
m_{1} & m_{2} & m_{3}\end{array}\right)\,.
\]

It should be mentioned that this recipe not only enables us to establish
the rotational invariance of any $N$-point correlation function,
but it also furnishes a straightforward definition of unbiased estimators
for the $N$-point functions. All we have to do is to drop
the ensemble average of the primed $a_{\ell m}$'s. So, for example,
for the 2- and 3-point functions above, the unbiased estimators are
given respectively by:
\begin{eqnarray*}
\widehat{C}_{\ell} & = & \frac{1}{2\ell+1}\sum_{m}a_{\ell m}a_{\ell m}^{*}\\
\widehat{B}_{\ell_{1}\ell_{2}\ell_{3}} & = & \sum_{m_{1},m_{2},m_{3}}a_{\ell_{1}m_{1}}a_{\ell_{2}m_{2}}a_{\ell_{3}m_{3}}
\left(\begin{array}{ccc}\ell_{1} & \ell_{2} & \ell_{3}\\
m_{1} & m_{2} & m_{3}\end{array}\right)\,.
\end{eqnarray*}

Notice that isotropy plays the same role, in $S^2$, that homogeneity plays 
in $\mathbb{R}^3$. What enforces homogeneity in $\mathbb{R}^3$ is
the Fourier-space $\delta$-functions, as in the discussion around
Fig. 1. However, in
$S^2$ the equivalent of the Fourier modes are the harmonic modes,
for which there is only a discrete notion of orthogonality -- and no
Dirac $\delta$-function. What we found
above is that the Wigner 3-j symbols play the same role as the
Fourier space $\delta$-functions: they are the enforcers of 
isotropy (rotational invariance) for the $N$-point angular correlation 
function. Hence, the diagrammatic representations of the constituents
of the $N$-point functions in Fourier (Fig. 1) and in harmonic space (Fig. 2)
really do convey the same physical idea -- one in $\mathbb{R}^3$,
the other in $S^2$.

\subsection{Gaussian and Statistically Anisotropic
models\label{sub:Gaussian-and-SA-models}}

In the last section, we have developed an algorithm which enables
one to establish the rotational invariance of any $N$-point correlation
function. As we have shown, this is also an algorithm for building
unbiased estimators of non-Gaussian correlations. In this section
we will change the perspective and analyze the case of Gaussian but
statistically anisotropic models of the Universe. 

There are many ways in which statistical anisotropy may be manifested
in CMB. From a fundamental perspective, a short phase of inflation
which produces just enough e-folds to solve the standard Big Bang
problems may leave imprints on the largest scales of the Universe,
provided that the spacetime is sufficiently anisotropic at the onset
of inflation \cite{Pitrou:2008gk}.
Another source of anisotropy may result from our inability to efficiently
clean foreground contaminations from temperature maps. Usually, the
cleaning procedure involves the application of a mask function in
order to eliminate contaminations of the galactic plane from raw data.
As a consequence, this procedure may either induce, as well as hide some
anomalies in CMB maps \cite{Abramo:2009fe}.

It is important to mention that these two examples can be perfectly
treated as Gaussian: in the first case, the anisotropy of the spacetime
can be established in the linear regime of perturbation theory, and
therefore will not destroy gaussianity of the quantum modes, provided
that they are initially Gaussian. In the second case, the mask acts
linearly over the temperature maps, therefore preserving its probability
distribution \cite{Hivon:2001jp}.

\subsubsection{Primordial anisotropy}

Recently, there have been many attempts to test the isotropy of the
primordial Universe through the signatures of an anisotropic inflationary
phase \cite{Pereira:2007yy,Pitrou:2008gk,Gumrukcuoglu:2007bx,Ackerman:2007nb,Shtanov:2009wp}.
A generic prediction of such models is the linear coupling of the
scalar, vector and tensor modes through the spatial shear, which is
in turn induced by anisotropy of the spacetime \cite{Pereira:2007yy}.
Whenever that happens, the matter power spectrum, defined in 
a similar way as in Eq. (\ref{Npfk}),
will acquire a directionality dependence due to this type of see-saw 
mechanism.
This dependence can be accommodated in a harmonic expansion of the
form:
\begin{equation}
P(\vec{k})=\sum_{\ell,m}r_{\ell m}(k)Y_{\ell m}(\hat{k}) \; ,
\label{aniso-power-spec}
\end{equation}
where the reality of $P(\vec{k})$ requires that 
$r_{\ell m}(k)=(-1)^{m}r_{\ell,-m}^{*}(k)$.
Given that temperature perturbations $\Theta(\vec{x})$ are real, their
Fourier components must satisfy the relation
$\Theta(\vec{k})=\Theta^{*}(-\vec{k})$.
This property taken together with the definition (\ref{Npfk})
implies that:
\begin{equation}
P(\vec{k})=P(-\vec{k}) \; ,
\label{Pkeven}
\end{equation}
which in turn restricts the $\ell$'s in (\ref{aniso-power-spec})
to even values. Also, note that by relaxing the assumption of
spatial isotropy, we are only breaking a continuous spacetime symmetry,
but discrete symmetries such as parity should still be present
in this class of models. Indeed, by imposing invariance of the spectrum 
under the transformation $z\rightarrow-z$, we find that $(-1)^{\ell-m}=1$.
Similarly, invariance under the transformations $x\rightarrow-x$ and $y\rightarrow-y$
imply the conditions $r_{\ell m}=(-1)^{m}r_{\ell,-m}$ and $r_{\ell m}=r_{\ell,-m}$,
respectively. Gathering all these constraints with the parity of
$\ell$, we conclude that 
\begin{equation}
r_{\ell m}\in\mathbb{R},\qquad\ell,m\in2\mathbb{N}\,.
\label{regras-selecao}
\end{equation}
That is, from the initial $2\ell+1$ degrees of freedom, only $\ell/2+1$
of them contribute to the anisotropic spectrum \cite{Pitrou:2008gk}.

\subsubsection{Signatures of statistical anisotropies}

The selection rules (\ref{regras-selecao}) are the most generic predictions
we can expect from models with global break of anisotropy. We will
now work out the consequences of these rules to the temperature power
spectrum, and check whether they can say something about the CMB large
scale anomalies. 

From the expressions (\ref{2pth}) and (\ref{aniso-power-spec}) we
can immediately calculate the most general anisotropic covariance matrix
\cite{Pullen:2007tu}:

\begin{equation}
\langle a_{\ell_{1}m_{1}}a_{\ell_{2}m_{2}}^{*}\rangle=\sum_{\ell_{3},m_{3}}
\mathcal{G}_{m_{1}m_{2}m_{3}}^{\ell_{1}\ell_{2}\ell_{3}}H_{\ell_{1}\ell_{2}}^{\ell_{3}m_{3}} \; ,
\label{aniso-cov-matrix}
\end{equation}
where:
\begin{equation}
\mathcal{G}_{m_{1}m_{2}m_{3}}^{\ell_{1}\ell_{2}\ell_{3}}=(-1)^{m_1}
\sqrt{\frac{(2\ell_{1}+1)(2\ell_{2}+1)(2\ell_{3}+1)}{4\pi}}\left(\begin{array}{ccc}
\ell_{1} & \ell_{2} & \ell_{3}\\
0 & 0 & 0\end{array}\right)\left(\begin{array}{ccc}
\ell_{1} & \ell_{2} & \ell_{3}\\
-m_{1} & m_{2} & m_{3}\end{array}\right) \, ,
\label{gaunt-coef}
\end{equation}
are the Gaunt coefficients resulting from the integral of three spherical
harmonics (see the Appendix). These coefficients are zero unless the
following conditions are met:
\begin{equation}
\begin{cases}
\ell_{1}+\ell_{2}+\ell_{3} & \in2\mathbb{N}\\
m_{1}+m_{2}+m_{3} & =0\\
|\ell_{i}-\ell_{j}|\leq\ell_{k}\leq\ell_{i}+\ell_{j} & \forall i,j,k\in\{1,2,3\} \; .
\end{cases}
\label{regras-selecao-2}
\end{equation}
The remaining coefficients in (\ref{aniso-cov-matrix}) are given
by: 
\[
H_{\ell_{1}\ell_{2}}^{\ell_{3}m_{3}}=4\pi
i^{\ell_{1}-\ell_{2}}\int_{0}^{\infty}{d\log k}\;
r_{\ell_{3}m_{3}}(k)j_{\ell_{1}}(kR)j_{\ell_{2}}(kR) \, ,
\]
and correspond to the
anisotropic generalization of temperature power spectrum (\ref{2pCl}).

The selection rules (\ref{regras-selecao-2}), taken together with
(\ref{regras-selecao}), lead to important signatures in the CMB. In
particular, since $\ell_{3}$ is even, the quantity $\ell_{1}\pm\ell_{2}$
must also be even, {\it i.e.}, multipoles with different parity do not
couple to each other in this class of models:

\begin{equation}
\langle a_{\ell_{1}m_{1}}a_{\ell_{2}m_{2}}^{*}\rangle=0\,,
\qquad\ell_{1}+\ell_{2}=\mbox{odd}\,.\label{even-even}
\end{equation}
This result is in fact expected on theoretical grounds, because by
breaking a continuous symmetry (isotropy) we cannot expect to generate
a discrete signature (parity) in CMB. However, notice that the absence of
correlations between, say, the quadrupole and the octupole, does not imply
that there will be no alignment between them. One example of this would be
a covariance matrix of the form $C_{\ell m} \delta_{\ell \ell'} \delta_{m
m'}$. If the $C_{\ell, 0}$ happen to be zero, for example, then all
multipoles will present a preferred direction (in this case, the $z$-axis.)


\subsubsection{Isotropic signature of statistical anisotropy}

We have just shown that a generic consequence of an early anisotropic
phase of the Universe is the generation of even-parity signatures
in CMB maps. Interestingly, these signatures may be present even in
the isotropic angular spectrum, since the $C_{\ell}$'s
acquire some additional modulations in the presence of statistical anisotropies. 
In principle, these modulations could be constrained by measuring an \textit{effective}
angular spectrum of the form
\begin{equation}
\langle a_{\ell m}a_{\ell m}^{*}\rangle=
C_{\ell}+\epsilon\sum_{\ell'>0}\mathcal{G}_{m,-m,0}^{\ell\ell\ell'}
H_{\ell\ell}^{\ell'0} \; ,
\label{Cl-efetivo} 
\end{equation}
where we have introduced a small $\epsilon$ parameter to quantify
the amount of primordial anisotropy. 

In order to constrain these modulations we have to build a statistical
estimator for $\langle a_{\ell m}a_{\ell m}^{*}\rangle$. Since we are looking for 
the diagonal entries of the matrix (\ref{aniso-cov-matrix}), a first
guess would be
\begin{equation}
\widehat{C}_{\ell}^{{\rm eff}}=\frac{1}{2\ell+1}
\sum_{m=-\ell}^{\ell}a_{\ell m}a_{\ell m}^{*}\,.
\label{Cl-efetivo-hat}
\end{equation}
To check whether this is an unbiased estimator of the effective angular spectrum,
we apply it to (\ref{aniso-cov-matrix}) and take its average:
\[
\langle C_{\ell}^{{\rm eff}}\rangle=C_{\ell}+\epsilon\sum_{\ell'>0,m}\mathcal{G}_{m,-m,0}^{\ell\ell\ell'}
H_{\ell\ell}^{\ell'0}\,.
\]
Using the definition (\ref{gaunt-coef}) of the Gaunt coefficients,
the $m$-summation in the expression above becomes \cite{Varshalovich:1988ye}
\[
\sum_{m}(-1)^{\ell-m}\left(\begin{array}{ccc}
\ell & \ell & \ell'\\
m & -m & 0\end{array}\right)=\sqrt{2\ell+1}\delta_{\ell',0}=0
\]
where the last equality follows because $\ell'>0$. Consequently we
conclude that
\[
\langle\widehat{C}_{\ell}^{{\rm eff}}\rangle=C_{\ell}\neq 
\langle a_{\ell m}a_{\ell m}^{*}\rangle\,.
\]
At first sight, this result may seem innocuous, showing only that
this is not an appropriate estimator for (\ref{Cl-efetivo}). Note however 
that (\ref{Cl-efetivo-hat}) is in fact the estimator of the angular spectrum
usually applied to CMB data under the assumption of statistical isotropy.
In other words, by means of the usual procedure we may be neglecting important 
information about statistical anisotropy. Moreover, the cosmic variance induced 
by the application of this estimator on anisotropic CMB maps is small, because, 
as it can easily be checked:
\[
\langle(\widehat{C}_{\ell}^{{\rm eff}}-C_{\ell})(\widehat{C}_{\ell'}^{{\rm eff}}-C_{\ell'})\rangle=\frac{2}{2\ell+1}\, C_{\ell}^{2}\,\delta_{\ell\ell'}+\mathcal{O}(\epsilon^{2}) \; .
\]

This result shows that the construction of statistical estimators
strongly depends on our prejudices about what non-gaussianity and
statistical anisotropy should look like. Consequently, an estimator
built to measure one particular property of the CMB may equally well
hide other important signatures. One possible solution to this problem 
is to let the construction
of our estimators be based on what the observations seem to tell us, 
as we will do in the next Section.



\newpage
\part{Cosmological Tests}

So much for mathematical formalism. We will now turn to the question of how
the hypotheses of gaussianity and statistical isotropy of the Universe can
be tested. Though we are primarily interested in applying these tests to
CMB temperature maps, most of the tools we shall be dealing with can be
applied to polarization ($E$- and $B$-mode maps) 
and to catalogs of large-scale structures as well.

Testing the gaussianity and SI of our Universe is a difficult task. 
Specially because, as we have seen, there is only one Gaussian and SI
Universe, but infinitely many universes which are neither Gaussian nor
isotropic. So what type of non-gaussianity and statistical anisotropy
should we test for? In order to attack this problem we can follow two
different routes. In the {\it bottom-up} approach, models for the cosmic
evolution are formulated in such a way as to account for some specific
deviations from gaussianity and SI. These physical principles range from
non-trivial cosmic topologies 
\cite{Luminet:2003dx,Riazuelo:2003ud,HipolitoRicaldi:2005eh}, primordial
magnetic fields
\cite{Kahniashvili:2008sh,Kahniashvili:2008hx,Bernui:2008ve,Caprini:2009vk,
Seshadri:2009sy}, 
local
\cite{Gordon:2005ai,Prunet:2004zy,Campanelli:2007qn,Campanelli:2009tk} and
global
\cite{Pereira:2007yy,Pitrou:2008gk,Gumrukcuoglu:2007bx,Ackerman:2007nb}
manifestation of anisotropy, to non-minimal inflationary models
\cite{Shtanov:2009wp,Erickcek:2008sm,Erickcek:2009at,Donoghue:2007ze,
Kawasaki:2008sn,Kawasaki:2008pa}. The main advantage of the bottom-up
approach is that we know exactly what feature of the CMB is being measured.
One of its drawbacks is the plethora of different models and mechanisms
that can be tested.

The second possibility is the {\it top-down}, or model-independent 
approach. Here, we are not concerned with the mechanisms responsible for
deviations of gaussianity or SI, but rather with the qualitative features
of any such deviation. Once these features are understood, we can use them
as a guide for model building. Examples here include constructs of {\it a
posteriori} statistics \cite{Bernui:2005pz,Abramo:2006gw,Hanson:2009gu}
and functional modifications of the two-point correlation function 
\cite{Abramo:2009fe,Pullen:2007tu,Hajian:2003qq,Hajian:2004zn,Hajian:2005jh,
Pereira:2009kg}. 

In the next section we will explore two different model-independent tests: 
one based on functional modifications of the two-point correlation
function, and another one based on the so-called Maxwell's multipole
vectors.

\section{Multipole vectors}\label{sec:mv}

Multipole vectors were first introduced in cosmology by Copi {\it et al.} 
\cite{Copi:2003kt} as a new mathematical representation for the primordial
temperature field, where each of its multipoles $\ell$ are represented by
$\ell$ unit real vectors. Later it was realized that this idea is in 
fact much older \cite{Weeks:2004cz}, being proposed originally by J. C.
Maxwell in his {\it Treatise on Electricity and Magnetism}. 

The power of this approach is that the multipole vectors can be entirely 
calculated in terms of a temperature map, without any reference to external
reference frames. This make them ideal tools to test the morphology of CMB
maps, like the quadrupole-octupole alignment. 

The purpose of the following presentation is only comprehensiveness. 
A mathematically rigorous introduction to the subject can be found in 
references  \cite{Katz:2004nj,Weeks:2004cz,EHobson}, as
well as other review articles in this review volume.

\subsection{Maxwell's representation of harmonic functions}

We start our presentation of the multipole vectors by recalling some 
terminology. A harmonic function in three dimensions is any (twice
differentiable) function $h$ that satisfies Laplace's equation:
\begin{equation}
\nabla^{2}h=0 \; ,
\label{laplace}
\end{equation}
where $\nabla^2$ is the Laplace operator. In spherical coordinates, the 
formal solution to  Laplace's equation which is regular at the origin
($r=0$) is:
\begin{equation}
h=\sum_{\ell=0}^{\infty}h_{\ell}(r,\theta,\varphi)\,,
\qquad
h_{\ell}=\sum_{m=-\ell}^\ell a_{\ell m}r^{\ell}Y_{\ell m}(\theta,\varphi)
\,.
\label{f-spherical}
\end{equation}
The functions $r^\ell Y_{\ell m}$ are known as the {\it solid spherical harmonics} 
\cite{EHobson}. Since they agree with the usual spherical harmonics on the unit sphere, 
it is sometimes stated in the literature that the latter form a set of harmonic functions. 
This is an abuse of nomenclature though, and the reader should be careful.

Given the scalar nature of Laplace's operator, it is possible to find 
solutions to Eq.(\ref{laplace}) in terms of Cartesian coordinates. Such
solutions can be constructed by combining homogeneous
polynomials\footnote{A homogeneous polynomial is a sum of monomials, all of
the same order.} of order $\ell$:
\begin{equation}
h=\sum_{\ell=0}^{\infty}h_{\ell}(x,y,z)\,,
\quad
h_{\ell}=\sum_{abc}^{\ell}\lambda_{abc}x^{a}y^{b}z^{c}\,,\quad(a+b+c=\ell)\,.
\label{f-cartesian}
\end{equation}

In three dimensions, the most general homogeneous polynomial of order 
$\ell$ contains $(\ell+2)!/(2!\ell!)$ independent coefficients. However,
since each polynomial must independently satisfy Eq. (\ref{laplace}),
precisely $\ell!/(2!(\ell-2)!)$ of these coefficients will depend on each
other. This constraint leaves us with 
$(\ell+1)(\ell+2)/2-\ell(\ell-1)/2=2\ell+1$ independent degrees of freedom
in each multipole $\ell$ -- which is, of course, the same number of
independent degrees of freedom appearing in Eq. (\ref{f-spherical}). 

Based on this analysis, Maxwell introduced his own representation of 
harmonic functions. He noticed that by successively applying directional
derivatives of the form $\vec{v}\cdot\nabla\equiv\nabla_{\vec{v}}$ over the
monopole potential $1/r$, where $r=\sqrt{x^2+y^2+z^2}$ and $\vec{v}$ is a
unit vector, he could construct solutions of the form (\ref{f-cartesian}).
That is:
\begin{equation}
f_\ell(x,y,z)=\lambda_\ell\left.\nabla_{\vec{v}_\ell}\dots
\nabla_{\vec{v}_2}\nabla_{\vec{v}_1}\frac{1}{r}\right|_{r=1} \; ,
\label{f-maxwell}
\end{equation}
where $\lambda_\ell$ are real constants. We are now going to show that 
this construction does indeed lead to solutions of the form
(\ref{f-cartesian}). First, note that there is a pattern which emerges 
from successive application of directional derivatives over the monopole
function:
\begin{eqnarray}
f_{0} & = & \frac{1}{r}\nonumber \\
f_{1} & = & \nabla_{\vec{v}_1}f_{0}=\frac{-\vec{v}_1\cdot\vec{r}}{r^{3}}
\nonumber \\
f_{2} & = & \nabla_{\vec{v}_2}f_{1}=\frac{3(\vec{v}_2\cdot\vec{r})(\vec{v}_1
\cdot\vec{r})-r^{2}(\vec{v}_1\cdot\vec{v}_2)}{r^{3}}  \; ,
\label{construcao-fl}
\end{eqnarray}
and so on. By induction, one can show that the general expression will be 
given by:
\begin{equation}
f_{\ell}=\frac{(-1)^{\ell}(2\ell-1)!!\prod_{i=1}^{\ell}\vec{v}_i
\cdot\vec{r}+r^{2}Q_{\ell-2}}{r^{2\ell+1}}\,,
\label{fl-produto}
\end{equation}
where $Q_{\ell-2}$ is a homogeneous polynomial of order $\ell-2$ which 
only involves combinations of the vectors $\vec{v}_i$ and $\vec{r}$.

The numerator of the function $f_\ell$ given by (\ref{fl-produto}) is 
clearly a homogeneous polynomial of order $\ell$ (as one can easily check
for some $\ell$'s and also prove by mathematical induction.) A not so
obvious result is that this polynomial is also harmonic. To prove that, 
let us define:
\begin{equation}
g_{\ell}=(-1)^{\ell}(2\ell-1)!!\prod_{i=1}^{\ell}\vec{v}_i\cdot\vec{r}+
r^{2}Q_{\ell-2} \, ,
\label{gl}
\end{equation}
and consider the application of the operator $\nabla^2$ over the 
combination $r^{\alpha}g_{\ell}$. For the $x$-component we get: 
\[
\partial_{x}^{2}(r^{\alpha}g_{\ell})=r^{\alpha}\partial_{x}^{2}g_{\ell}
+2\alpha r^{\alpha-2}x\partial_{x}g_{\ell}+[\alpha r^{\alpha-2}+\alpha(\alpha-2)x^{2}r^{\alpha-4}]g_{\ell}\,.
\]
Repeating this process for the $y$- and $z$-components and then adding the 
results, we find: 
\begin{eqnarray*}
\nabla^{2}(r^{\alpha}g_{\ell}) & = & r^{\alpha}\nabla^{2}g_{\ell}+2\alpha 
r^{\alpha-2}(x\partial_{x}g_{\ell}+y\partial_{y}g_{\ell}+
z\partial_{z}g_{\ell})
+\alpha(\alpha+1)r^{\alpha-2}g_{\ell}\\
& = &r^{\alpha}\nabla^{2}g_{\ell}+\alpha(\alpha+2\ell+1)
(r^{\alpha-2}g_{\ell})\,,
\end{eqnarray*}
where in the last step we have used Euler's theorem on homogeneous
functions, {\it i.e.}, $\vec{r} \cdot \nabla g_\ell = \ell g_\ell$.
If we now choose $\alpha=-(2\ell+1)$, we find immediately that
\[
\nabla^{2}\left(\frac{g_{\ell}}{r^{2\ell+1}}\right)=
\frac{\nabla^{2}g_{\ell}}{r^{2\ell+1}}\,.
\]

By construction, the left-hand side of the above expression is equal to 
$\nabla^2 f_\ell$. But according to the definition (\ref{f-maxwell}) this
quantity is also zero, since Laplace's operator commutes with directional
derivatives and $\nabla^2 (1/r)=0$ for $r>0$. Therefore, $\nabla^2 g_\ell=0$ and
$g_\ell$ is harmonic, which completes our proof.

In conclusion, Maxwell's construction of harmonic functions, 
Eq. (\ref{f-maxwell}), is completely equivalent to the standard
representation in terms of spherical harmonics. More importantly, this
gives a one-to-one relationship between temperature maps (given by the
$a_{\ell m}$'s) and $\ell$ unit vectors $\vec{v}_i$. This means that the
multipole vectors can be directly calculated from a CMB map, without any
reference to external reference frames or additional geometrical
constructs. The reader interested in algorithms to construct the vectors
from CMB maps may check references \cite{Copi:2003kt,Weeks:2004cz}, as well
as the other articles in this volume that review this approach.

\subsection{Multipole vectors statistics}

It should be clear from the discussion above that the multipole
vectors give an intuitive way to discover, interpret and visualize
phase correlations between different multipoles in the CMB maps.
But they also can reveal intra-multipole features, such as planarity
(when a given multipole presents a preferred plane.)
Some of the most conspicuous hints of statistical anisotropy
in the CMB, like the quadrupole-octupole alignment, have indeed 
been first found with less mathematically elegant methods 
\cite{deOliveiraCosta:2003pu,Tegmark:2003ve}, 
but are best described in terms of the multipole
vector formalism \cite{Copi:2003kt,Schwarz:2004gk,Copi:2005ff,Weeks:2004cz}.
However, this case should also sound an alarm, because
some feature of a (presumably) random realization was found, then
further scrutinized with a certain test which was, to some extent (intentionally
or not) tailored to single out that very feature.

The pitfalls of aprioristic approaches are sometimes unavoidable,
and all we can do is take a second look at our sample with a more
generic set of tools, to try to assess how significant our result
really is in the context of a larger set of statistical tests.
Multipole vectors are, in fact, ideally suited to this, since they
can be found for all multipoles, and it is relatively easy to construct
scalar combinations of these vectors with the usual methods
of linear algebra. Also convenient is the fact that simulating
these vectors from maps, or even directly, is also relatively easy, so
the standard model of a Gaussian random field can be easily
translated into the pdf's for the tests constructed with the
multipole vectors.
An important drawback of the multipole vectors is that,
because they are computed in terms of equations which are non-linear in 
the temperature fields, the distribution functions
of statistical tests involving these vectors are highly non-Gaussian \cite{Dennis:2007jk}.
This means that these statistics usually have to be estimated by means of simulations
(usually assuming that the underlying temperature field is itself
Gaussian.) Another delicate issue is how stable the multipole vectors
are to instrumental noise and sky cut -- and here, again, we must
rely on numerical simulations to compare the observations
with theoretical models.

In order to see how one should go about constructing a general
set of tests, it should be noted first that the multipole vectors define
{\it directions}, but they have no sense (they are ``headless vectors''.)
This follows from the fact that the sign of the constants 
$\lambda_\ell$ are degenerate with the sign of the vectors 
$\vec{v}_{\ell, p}$ ($p=1,\ldots, \ell$ .) 
Hence, the first requirement is that our tests
should be independent of this sign ambiguity.
Notice that this ambiguity extends to ancillary constructs such as the
normal vectors, defined as the vector product:
\be
\label{normvec}
\vec{w}_{\ell, p p'} \equiv \vec{v}_{\ell, p} \otimes \vec{v}_{\ell, p'} \; .
\ee
Mathematically, the sign ambiguity implies that the multipole vectors
belong to the quotient space $S^2/\mathbb{Z}_2$ (also known as $\mathbb{R} P^2$), 
while the normal vectors belong to $\mathbb{R}^3/\mathbb{Z}_2$.

In principle, any positive-definite scalar constructed through
the multipole or the normal vectors is ``fair game'', but there
are some guidelines, e.g., one should avoid double-counting 
the same degrees of freedom. Below we review some of these 
tests, based on the work of Ref. \cite{Abramo:2006gw}.

\subsubsection{The $R$ statistic}

The first example of a test involving the multipole vectors would
be a scalar product, such as $\vec{v} \cdot \vec{v} \, '$. The most
natural test would involve asking whether the $\ell$ multipole vectors
of the given multipole $\ell$ are especially aligned or not. This
means computing:
\be
\label{R_stat}
R_{\ell \ell} = \frac{2}{\ell (\ell-1)} \sum_{p, p'>p}^\ell | \vec{v}_{\ell, p} \cdot \vec{v}_{\ell, p'} | \; ,
\ee
where the normalization was introduced 
to make $0 \leq R_{\ell \ell} \leq 1$.

This idea could be generalized to test alignments between multipole
vectors at different multipoles:
\be
\label{R_stat2}
R_{\ell \ell'} = \frac{1}{\ell \ell'} \sum_{p, p'} | \vec{v}_{\ell, p} \cdot \vec{v}_{\ell', p'} | \; .
\ee
In fact, the quadrupole-octupole alignment can already be seen with this simple
test: for essentially all CMB maps the significance of the alignment as measured
by $R_{23}$ is of the order of 90-95\% C.L. \cite{Abramo:2006gw}.

\subsubsection{The $S$ statistic}

The second most natural test does not involve directly the multipole vectors
themselves, but the normal vectors that can be produced by taking
the vector product between the multipole vectors. So, we take:
\be
\label{nv}
\vec{w}_{\ell, p p'} = \vec{v}_{\ell, p} \otimes \vec{v}_{\ell', p'} \; .
\ee
Notice that the number of normal vectors for a given multipole
$\ell$ is $l=\ell(\ell-1)/2$ -- so the number of normal vectors grows
rapidly for larger multipoles, making it harder to use and
meaning that the same degrees of freedom may be overcounted,
at least for $\ell > 3$.

Again, the best strategy is simplicity: we can ask whether the normal
vectors are aligned, within and between multipoles.
This means computing:
\be
\label{S_stat}
S_{\ell \ell} = \frac{2}{l (l-1)} \sum_{p, p'>p}^l | \vec{w}_{\ell, p} \cdot \vec{w}_{\ell, p'} | \; ,
\ee
where the normalization was introduced again to make  $0 \leq S_{\ell \ell} \leq 1$.

And yet again this idea is easily generalized to test alignments between normal
vectors at different multipoles:
\be
\label{S_stat2}
S_{\ell \ell'} = \frac{1}{l l'} \sum_{p, p'} | \vec{w}_{\ell, p} \cdot \vec{w}_{\ell', p'} | \; .
\ee
With this test, the quadrupole-octupole alignment is much more significant:
for essentially all CMB maps the test $S_{23}$ 
deviates from the expected range with 98\% C.L. \cite{Abramo:2006gw}.

\subsubsection{Other tests with multipole vectors}

One can go on and expand the types of tests using both multipole
and normal vectors. One idea would be, e.g., to disregard the moduli
of the normal vectors:
\be
\label{D_stat}
D_{\ell \ell'} = \frac{1}{l l'} \sum_{p, p'} | \hat{w}_{\ell, p} \cdot \hat{w}_{\ell', p'} | \; .
\ee
This test is therefore insensitive to the relative angle between the multipole
vectors that produce any given normal vectors. This idea is similar
to the planar modulations that will be discussed in the next Section.
With this test, the quadrupole-octupole alignment is significant to
about 95-98\% C.L. This test can also be generalized to a self-alignment
test ($\ell=\ell'$), with just an adjustment to the normalization.

Another possibility would be to measure the alignment between multipole
vectors and normal vectors:
\be
\label{B_stat}
B_{\ell \ell'} = \frac{1}{\ell l'} \sum_{p, p'} | \vec{v}_{\ell, p} \cdot \vec{w}_{\ell', p'} | \; .
\ee
Of course, this test cannot be easily generalized to a self-alignment.
With the $B_{23}$ test, the quadrupole-octupole alignment is significant to
about 95-98\% C.L.

We could go on here, but it should be clear that all the information
(the $2\ell +1$ real degrees of freedom for each multipole $\ell$)
has already been exhausted in the tests above.

\section{Temperature correlation function}\label{sec:tcf}

Despite their strong cosmological appeal, the multipole vectors have some 
limitations. Not only their directions in the CMB sky are sometimes 
difficult to interpret physically \cite{Copi:2005ff}, they also have the
additional drawback of mixing, in a non-trivial manner, information on both
gaussianity and SI of the map being analyzed
\cite{Copi:2005ff,Dennis:2007jk}. 

Another way of quantifying deviations in the standard statistical 
framework of cosmology is through functional modifications of the two-point
correlation function \cite{Pullen:2007tu,Hajian:2003qq}. Although this
approach does not offer an optimal separation between gaussianity and SI
(which is, by the way, an open problem in this field), working with the
two-point correlation function makes it easier to test Gaussian models of
statistical anisotropy \cite{Pereira:2009kg,Abramo:2009fe}.\\

The most general 2-point correlation function (2pcf) of two independent 
unit vectors is a function $C$ of the form
\[
C:S^2\times S^2\rightarrow\mathbb{R}\,.
\]
We have seen in \S\ref{sec:statistics-harm-space} that, if we choose 
spherical coordinates $(\theta_i,\varphi_i)$ to describe each vector
$\hat{n}_i$, the function above can be decomposed either in terms of two
spherical harmonics $Y_{\ell_im_i}$ or in terms of the bipolar spherical 
harmonic $\mathcal{Y}^{LM}_{\ell_1\ell_2}$. In any case, the 2pcf will have
the following functional dependence
\begin{equation}
C=C(\theta_1,\varphi_1,\theta_2,\varphi_2)\,.
\label{C-Souradeep}
\end{equation}

\begin{figure}[H]
\begin{centering}
\includegraphics[clip,scale=0.4]{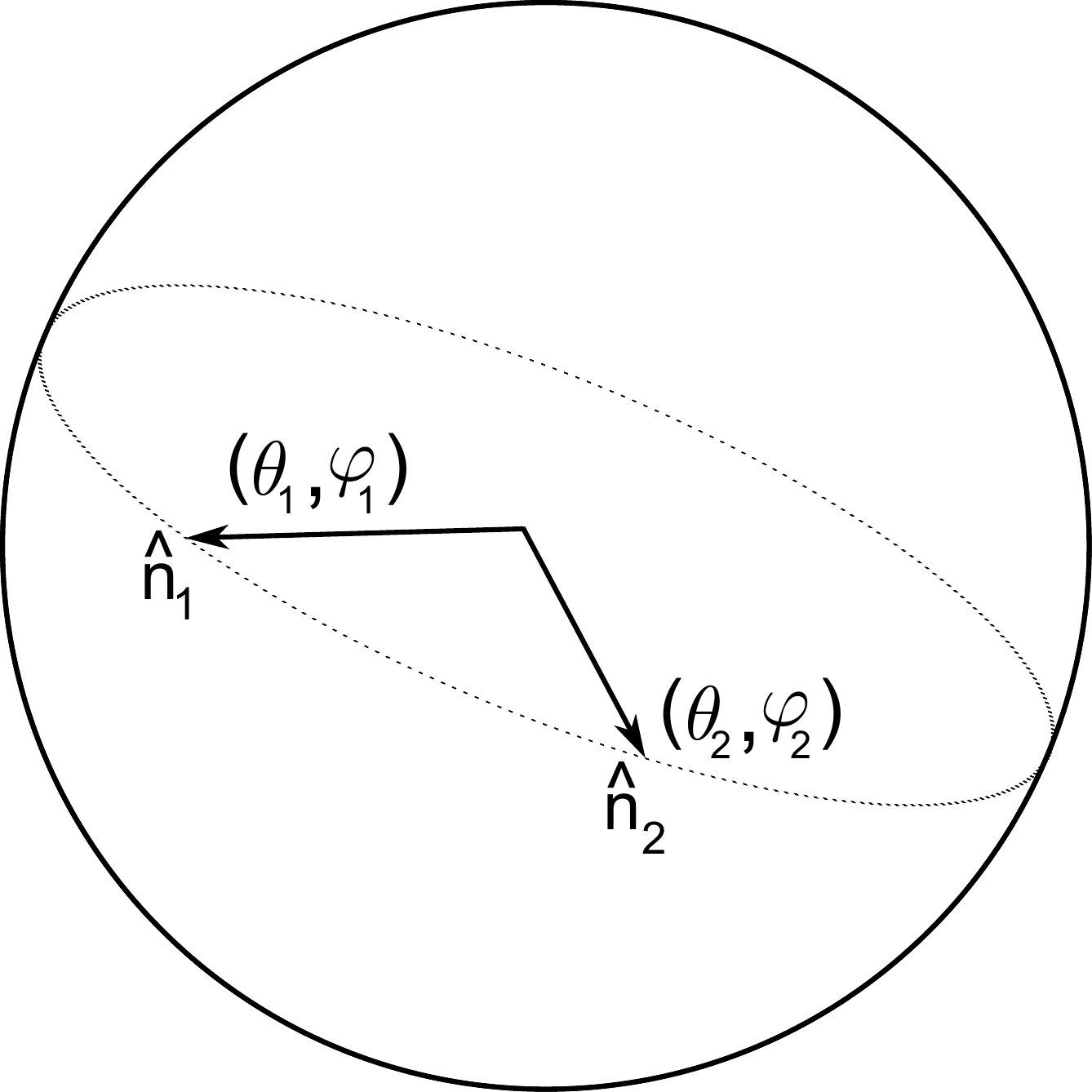} 
\label{sphere1}
\par\end{centering}
\caption{Geometrical representation of the 2pcf in terms of two unit 
vectors.}
\end{figure}

This function is absolutely general. If the Universe has any cosmological 
deviation of isotropy, whatever it is, it can be described by the function
above (see also Fig.(\ref{sphere1})). 

Unfortunately, this function will be of limited theoretical interest 
unless we have some hints on how to select its relevant degrees of freedom.
This difficulty is in fact a general characteristic of model-independent
tools, which at some stages forces us to rely on our 
theoretical prejudices about the statistical nature of the Universe in
order to construct estimators of non-gaussianity and/or statistical
anisotropy. Nonetheless, it is still possible to construct statistical
estimators of anisotropy based on Eq.(\ref{C-Souradeep}). For example, 
Hajian and Souradeep \cite{Hajian:2003qq} have constructed an unbiased
estimator $\kappa_\ell$ for this function based solely on the requirement
that this estimator should be rotationally invariant. Although it is true
that any statistically significant $\kappa_\ell>0$ will point 
towards anisotropy, it is not clear what type of anisotropy is being
detected by this estimator. 
\\

We can still use Eq. (\ref{C-Souradeep}) to search for deviations of SI if
we restrict its domain to a smaller and non-trivial sub-domain. For
example, we can take the vectors $\hat{n}_1$ and $\hat{n}_2$ to be the
same, and expand a function of the form
\begin{equation}
C=C(\theta_1,\varphi_1)\,.
\label{C-Kamionkowski}
\end{equation}
which is equivalent to $C:S^2\rightarrow\mathbb{R}$. This form of the 2pcf 
makes it ideal for searching for power multipole moments in CMB, once a
suitable estimator is defined \cite{Pullen:2007tu}. Unfortunately, when 
we take $\hat{n}_1=\hat{n}_2$ we are in fact considering a one-point
correlation function, which by construction does not allows us to 
measure correlations between different points in the sky.\\

It seems in principle that the functions (\ref{C-Kamionkowski}) and 
(\ref{C-Souradeep}) are the only possibilities besides the isotropic 2pcf
Eq.(\ref{C-theta}). If not, what other combinations of the vectors
$\hat{n}_1$ and $\hat{n}_2$ can we consider? As a matter of fact, these two
vectors are geometrical quantities intuitively bound to our notion of
two-point correlation functions on the sphere. From this perspective they
are not fundamental quantities. In fact, we can equally well represent the
2pcf by a disc living inside the unit sphere, as shown in Fig. 4
\begin{figure}[H]
\begin{centering}
\includegraphics[clip,scale=0.4]{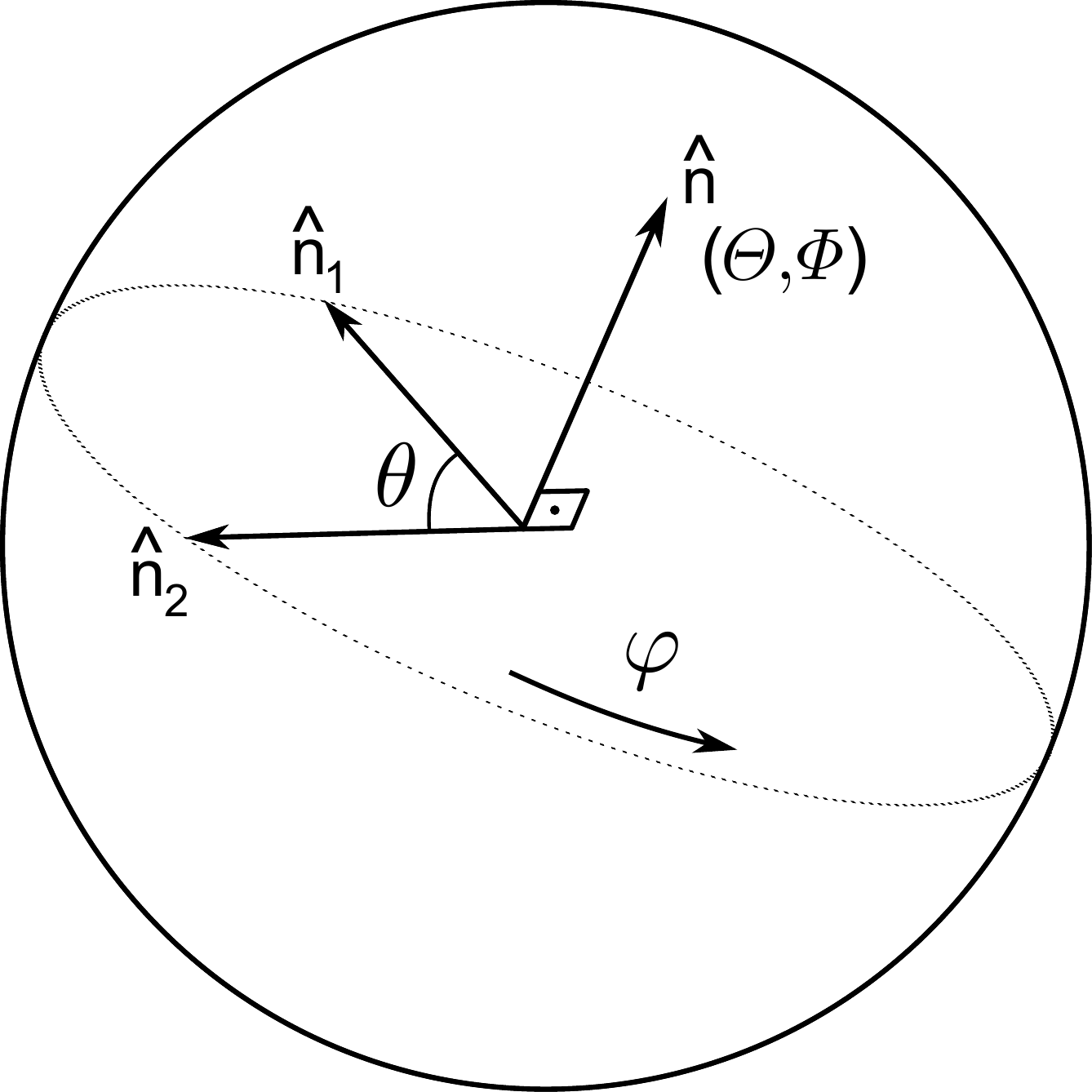}
\label{plane-sphere} 
\par\end{centering}
\caption{Geometrical representation of the 2pcf in terms of a unit disc (or plane).}
\end{figure}

In this representation, $\theta$ is the angle between the vectors 
$\hat{n}_1$ and $\hat{n}_2$, as usual. The normal to the plane, $\hat{n}$,
is represented by two spherical angles $(\Theta,\Phi)$. Finally, there is
an overall orientation $\varphi$ of the disc around its unit vector which
completes the four degrees of freedom contained in Eq.(\ref{C-Souradeep}).
We have found therefore another valid geometrical representation of the
most general 2pcf:
\begin{equation}
C=C(\Theta,\Phi,\theta,\varphi)
\label{C-APS}
\end{equation}

The main advantage of the above representation when compared to 
(\ref{C-Souradeep}) is its straightforward geometrical interpretation.
First, note that the angular separation $\theta$ of the isotropic 2pcf is
trivially included in this definition, and do not need to be obtained 
as a consequence of rotational invariance (see the discussion in
\S\ref{sec:Invariant-N-point}.) Second, by characterizing the correlation
function in terms of the geometrical components of a disc, we know exactly
what are the degrees of freedom involved. This makes it easier to construct
estimators of statistical anisotropy, alleviating the drawbacks of
model-independent approaches mentioned above. 

\subsection{Anisotropy through planarity}\label{sec:angular-planar}

An immediate application of the representation (\ref{C-APS}) is its use in 
the search for planar deviations of isotropy in CMB
\cite{Pereira:2009kg,Abramo:2009fe}. Planar modulations of astrophysical
origin may play an important role to the CMB morphology. One example is the
role played by the galactic and ecliptic plane in the
quadrupole-octupole/north-south anomalies \cite{Copi:2005ff}. Also, it is
well known that our galactic plane is sensible source of foreground
contamination in the construction of cleaned CMB maps. These hints indicate
that CMB modulations induced by the disc in Fig. 4 is not
only a mathematical possibility, but perhaps also a symmetry of
cosmological relevance.

Since we are primarily interested in measuring planar modulations of CMB, 
but including the usual angular modulation as the isotropic limit of the
2pcf, we can consider only the azimuthal average of Eq. (\ref{C-APS}):
\begin{equation}
C\rightarrow\frac{1}{2\pi}\int_0^{2\pi}C(\Theta,\Phi,\theta,\varphi)
d\varphi\,.
\label{azimuthal-average}
\end{equation}
The resulting function can be easily expand in terms of simple special 
functions as
\begin{equation}
C(\Theta,\Phi,\theta)=\sum_{\ell}\sum_{l,m}\frac{2\ell+1}{\sqrt{4\pi}}
\mathcal{C}_{\ell}^{lm}P_{\ell}(\cos\theta)Y_{lm}(\Theta,\Phi)\,,
\quad l\in2\mathbb{N}\,
\label{fc-aniso}
\end{equation}
where the restriction on the $l$-mode results from the symmetry 
$\hat{n}_1\leftrightarrow \hat{n}_2$ \cite{Yuri}.
The multipolar coefficients $\mathcal{C}_{\ell}^{lm}$ correspond to a
generalization of the usual angular power spectrum $C_{\ell}$'s. In fact,
they can be seen as the coefficients of a spherical harmonic decomposition
of the function $C_\ell(\hat{n})$, provided that this function suffers
modulations as we sweep planes on the sphere.

\subsubsection{Angular-planar power spectrum}

Since we are restricting our analysis to the Gaussian framework, the set 
of coefficients $\mathcal{C}_{\ell}^{lm}$ is all we need to characterize
the two-point correlation function. However, the final product of CMB
observations are temperature maps, and not correlation maps. What we need
then is an algebraic relation between the multipolar coefficients
$\mathcal{C}_{\ell}^{lm}$ and the temperature coefficients $a_{\ell m}$
defined in Eq.(\ref{DeltaT}). At first sight, this relation could be
obtained by equating expression (\ref{fc-aniso}) to its standard definition
in Eq. (\ref{APS}), and then using the orthogonality of the special
functions to isolate the $\mathcal{C}^{lm}_\ell$'s in terms of the
$a_{\ell m}$'s. But for that to work we need the relation between the set
of angles $(\Theta,\Phi,\theta)$ and
$(\theta_1,\varphi_1,\theta_2,\varphi_2)$ which, depending on the reference
frame we choose, is extremely complicated. Fortunately, all we need is the
relation:
\begin{equation}
\hat{n}_{1}\cdot\hat{n}_{2}=\cos\theta=\cos\theta_{1}\cos\theta_{2}+
\sin\theta_{1}\sin\theta_{2}\cos(\varphi_{1}-\varphi_{2})\,,
\label{cos-theta}
\end{equation}
together with a suitable choice of our coordinate system. For example, we
can use the invariance of the scalar product $\hat{n}_{1}\cdot\hat{n}_{2}$
and choose our coordinate system such that the disc of Fig. 4 lie in the
$xy$ plane. With this choice we will have:
\[
(\Theta,\Phi)=(0,0)\,,\qquad \cos\theta=\cos(\varphi_1-\varphi_2)\,,
\] 
and the integration over $\theta$ becomes simple. Once this is done, we
make a passive rotation of the coordinate system and then we integrate over
the remaining angles $\Theta$ and $\Phi$, which will then be given
precisely by the Euler angles used in the rotation. The details are rather
technical and can be found in the Appendix. The final expression is: 
\begin{equation}
\frac{(-1)^m\mathcal{C}_{\ell}^{lm}}{\sqrt{2l+1}}=2\pi\sum_{\ell_{1},m_{1}}
\sum_{\ell_{2},m_{2}}
\langle a_{\ell_{1}m_{2}}a_{\ell_{2}m_{2}}\rangle
\left(\begin{array}{ccc}l & \ell_{1} & \ell_{2}\\
-m & m_{1} & m_{2}\end{array}\right)I_{\ell_{1}\ell_{2}}^{l,\ell},
\label{Clmell}
\end{equation}
where
\begin{equation}
I_{\ell_{1}\ell_{2}}^{l,\ell}\equiv\sum_{m}(-1)^{m}\lambda_{\ell_{1}m}
\lambda_{\ell_{2}m}\left(\begin{array}{ccc}
l & \ell_{1} & \ell_{2}\\
0 & m & -m\end{array}\right)\int_0^\pi d(-\cos\theta)\, 
P_{\ell}(\cos\theta)e^{im\theta}\,,
\label{Int-l-ell}
\end{equation}
and where the $\lambda_{\ell_{i}m}$'s form a set of coefficients resulting
from the $\theta$ integration, which are zero unless $\ell_{i}+m=$ even
(see the Appendix for more details).

Expression (\ref{Clmell}) is what we were looking for. With this relation, 
the angular-planar power spectrum $\mathcal{C}^{lm}_\ell$ can be calculated
from first principles for any model predicting a specific covariance
matrix. Moreover, since the angular-planar function (\ref{fc-aniso}) is, after all, 
a correlation function, it should be possible to relate the angular-planar power spectrum $\mathcal{C}^{lm}_\ell$ to the bipolar power spectrum $\mathcal{A}^{LM}_{\ell_1\ell_2}$ 
of Hajian and Souradeep. In fact, by inverting expression (\ref{bips}) and plugging the 
result in (\ref{Clmell}), we find a linear relation between these two set of
coefficients:
\[
\mathcal{C}^{lm}_\ell=2\pi\sum_{\ell_1,\ell_2}A^{lm}_{\ell_1 \ell_2}
(-1)^{\ell_1-\ell_2}(2l+1)\,I^{\;l,\ell}_{\ell_1\ell_2}\,.
\]
Here, the set of geometrical coefficients $I^{\;l,\ell}_{\ell_1\ell_2}$ plays a similar 
role to the 3-j symbols in expression (\ref{bips}). Note also the angular-planar power 
spectrum has only three free indices, while the bipolar power spectrum has four. This 
is a consequence of the azimuthal average we took in (\ref{azimuthal-average}), which 
further constrains the degrees of freedom of the correlation function.

\subsubsection{Statistical estimators and $\chi^2$ analysis}

We have shown that the angular-planar power spectrum (\ref{Clmell}) is 
given in terms of an ensemble average of temperature maps. Evidently, we
cannot calculate it directly from data, for we have only one CMB map (the
one taken from our own Universe.) The best we can do is to estimate the
statistical properties of (\ref{Clmell}), like its mean and variance, and
see whether these quantities agree, in the statistical sense, with what we
would expect to obtain from a particular model of the Universe. 

The reader should note that this procedure is not new -- its limitation 
is due to the same cosmic variance which lead us to construct an estimator
for the angular power spectrum $C_\ell$ (see the discussion of
\S\ref{sec:estimators-and-cv}). For the same reason, we will need to
construct an {\it estimator} for the angular-planar power spectrum. An
obvious choice is:
\begin{equation}
\widehat{\mathcal{C}}^{\;lm}_\ell\equiv\;2\pi\sqrt{2l+1}\sum_{\ell_{1},m_{1}}
\sum_{\ell_{2},m_{2}}a_{\ell_{1}m_{1}}
a_{\ell_{2}m_{2}}\left(\begin{array}{ccc}l & \ell_{1} & \ell_{2}\\
m & m_{1} & m_{2}\end{array}\right)I_{\ell_{1}\ell_{2}}^{l,\ell} \; ,
\label{Clmell-hat}
\end{equation}
for in this case we have an unbiased estimator,
\[
\langle\widehat{\mathcal{C}}^{\;lm}_\ell\rangle=\mathcal{C}^{lm}_\ell, 
\]
regardless of the underlying model.

If we now have a model predicting the angular-planar power spectrum, we can
ask how good this model fit the observational data once it is calculated
using (\ref{Clmell-hat}). All we need is a simple chi-square ($\chi^2$)
goodness-of-fit test, which in our case can be written in the following
generalized form:
\begin{equation}
(\chi_{\nu}^{2})_{\ell}^{l}\equiv\frac{1}{2l+1}\sum_{m=-l}^{l}
\frac{|\widehat{\mathcal{C}}_{\ell}^{\;lm}-\mathcal{C}_{\ell}^{lm}|^{2}}
{(\sigma_{\ell}^{lm})^{2}} \; ,
\label{chi2}
\end{equation}
in which $\ell$ and $l$ are the angular and planar degrees of freedom, 
respectively, and where $\sigma^{lm}_\ell$ is just the standard deviation
of the estimator $\widehat{\mathcal{C}}^{\;lm}_{\ell}$. The $(2l+1)^{-1}$
factor accounts for the $2l+1$ planar degrees of freedom, and was
introduced for latter convenience. 

In \S\ref{sec:chi2-test} we will apply this test to the 5-year WMAP 
temperature maps in order to check the robustness of the $\Lambda$CDM model
against the hypothesis of SI. Before that, we shall stop and digress a
little about how observational uncertainties should be included in our
analysis.

\newpage
\section{Theory {\it v.} Observations: cosmic and statistical
variances}\label{sec:probabilities}

Until now we have been concerned with the formal aspects of non-Gaussian 
and statistically anisotropic universes, and how model-independent tests
might be designed to detect deviations of either gaussianity or statistical
isotropy. We will now discuss how such tests can be carried out and
interpreted once we possess cosmological data.

In a great variety of tests, statistical tools are designed to detect 
particular deviations of gaussianity or SI from cosmological data
like, for example, the CMB. The final outcome of these tests is usually a
probability (a pure number), which should be interpreted as the chance 
that a Universe like ours might result from an ensemble of ``equally 
prepared'' Gaussian and SI universes. An anomaly in CMB is usually
understood as a measure of how unlikely is a particular feature of our
Universe according to this specific test. The multipole vector statistics,
for example, when applied to a large number of simulated (Gaussian and SI)
CMB maps, show that only $\sim 0.01\%$ of these maps have a
quadrupole-octupole alignment as strong as WMAP maps
\cite{Copi:2005ff,Copi:2003kt}. 

There are two points to keep in mind when carrying this type of analysis. 
The first is that a particular ``detection'' may always turn out to be a
statistical fluctuation revealed by one specific tool. The robustness of an
anomaly then depend on the number of independent tests pointing to the same
result. The second is that the implementation of statistical tools is
sensitive to the way we extract information from data, requiring an
accurate separation between cosmological signal and
astrophysical/instrumental noise. \\

In this section we present a critical review of the standard procedure 
used to implement cosmological tests. We show that it does not account for
the intrinsic uncertainties of cosmological observations, which may
possibly lead to an under/over estimation of anomalies. We then present a
generalization of this process which naturally accounts for these
uncertainties.

\subsection{Standard Calculations}\label{sec:standard-calc}

Suppose $x$ is a random variable predicted by a particular
model\footnote{Rigorously, $x$ is only {\it one realization} of a random
variable $X$, which is a real-valued function defined on a sample space. 
By the same reason we should not call the $a_{\ell m}$'s in Eq.
(\ref{DeltaT}) random variables, though we shall stick to this nomenclature
throughout this text.} and that cosmological observations of this quantity
returned the value $x_0$. We would like to calculate the probability, {\it
according to this model}, that in a random Universe we would have $x\leq
x_0$. Assuming that $\mathcal{P}_{\scriptsize{\rm th}}$ is the (normalized)
probability density function (pdf) of $x$, this probability is commonly
defined to be 
\begin{equation}
P_{\leq}\equiv\int_{-\infty}^{x_{0}}\mathcal{P}_{\scriptsize{\rm th}}(x)\,
dx\,.
\label{P>0}
\end{equation}
The probability of having $x>x_{0}$ is then simply given by 
$P_{>}=1-P_{\leq}$. See the figure below.
\begin{center}
\begin{figure}[h]
\begin{centering}
\includegraphics[scale=1]{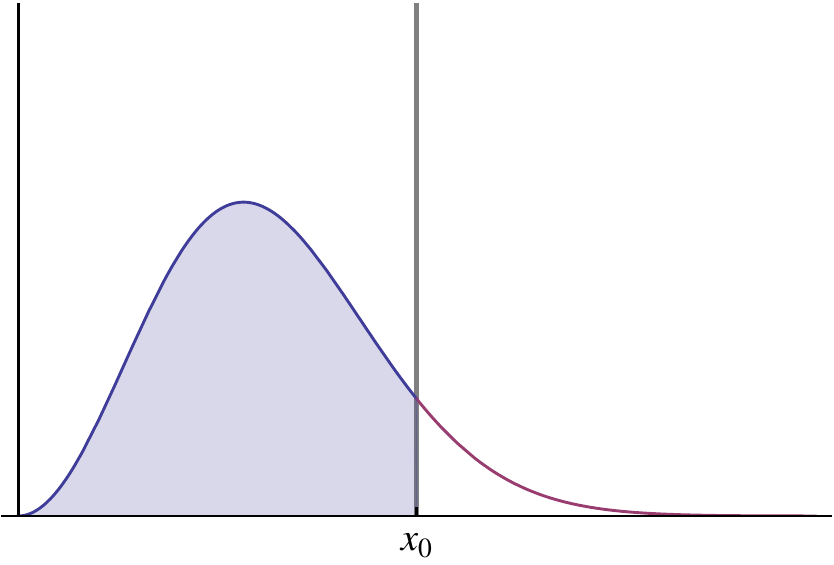}
\par\end{centering}
\caption{Probability density function for the theoretical variable $x$. The
shaded are is the probability Eq. (\ref{P>0}).}
\end{figure}
\par\end{center}

If $P_{\leq}$ is found to be too small or, equivalently, too high, we 
might be tempted to interpret $x_0$ as ``anomalous'' according to this
model. However, this definition of probability assumes that $x_0$ was
measured with infinite precision, and so it says nothing about an important
question we must deal with: typically, the measurement of $x_0$ has an
uncertainty which needs to be folded into the final probabilities that the
observations match the theoretical expectations.

Moreover, since no two equally prepared experiments will ever return the 
same value $x_0$, our measurements should also be regarded as random
events. In a more rigorous approach, we would have to consider $x_{0}$
itself as one realization of a random variable, conditioned to the distribution of the signal. 
In the case of CMB, however, this would be only part of the whole picture, since the
randomness of the measurements of $x_{0}$ should also be related to the way
this data is reduced to its final form. This happens because different
map cleaning procedures will lead to slightly different values for $x_{0}$.
This difference induces a variance in the data which reflects the remaining
foreground contamination of the temperature map. We will elaborate more on
this point through a concrete example, after we show how Eq. (\ref{P>0})
may be changed in order to include the indeterminacy of cosmological
measurements.

\subsection{Convolving probabilities}\label{sec:conv-prob}

The question we want to answer is: how to calculate the probability of $x$ 
being smaller than our measurements when the latter are also random events?
Let us suppose that our measurement is described by the random variable $y$
and that $x_0$ is its most probable value. The probability of having
$x\leq y$ is simply the probability of having
\[
z\equiv x-y \leq 0
\]
for some particular realization of the variable $y$. It should be clear by 
now that if we know the pdf of $z$, the probability we are looking for is
simply the area under this distribution for $-\infty\leq z \leq0$. The probability $P_\leq$ 
of $z$ being smaller or equal to zero can be calculated as:
\[
P_\leq\equiv P\{(x,y)|x-y\leq0\}=\iintop_{x-y\,\leq0}\mathcal{P}(x,y)\,dx\,dy\,.
\]
Now, under the hypothesis of independence of $x$ and $y$ we have 
$\mathcal{P}(x,y)=\mathcal{P}_{\scriptsize{\mbox{th}}}(x)\mathcal{P}_{\scriptsize{\mbox{obs}}}(y)$, 
where $\mathcal{P}_{\scriptsize{\mbox{obs}}}$ is the (normalized) probability distribution function
of the variable $y$. We can therefore rewrite the last expression as
\[
P_\leq=\int_{-\infty}^{\infty}\left[\int_{-\infty}^{y}\mathcal{P}_{\scriptsize{\mbox{th}}}(x)\,dx\right]
\mathcal{P}_{\scriptsize{\mbox{obs}}}(y)\,dy\,.
\]
If we now hold $y$ fixed and do the change of variables $x=u+y$, we get
\[
P_\leq=\int_{-\infty}^{0}\left[\int_{-\infty}^{\infty}\mathcal{P}_{\scriptsize{\mbox{th}}}(u+y)
\mathcal{P}_{\scriptsize{\mbox{obs}}}(y)\,dy\right]\,du\,.
\]
where we have changed the position of the integrals. The reader will now notice that term inside brackets is 
precisely the pdf we were looking for. Since there is nothing special about the variable $y$, we can equally well 
hold $x$ fixed and repeat the calculus, obtaining the symmetric version of this result. In fact, the final pdf for $z$ is 
nothing else than the convolution of the pdf's of each variable \cite{Ash}:
\begin{eqnarray}
\mathcal{P}(z)\equiv\left(\mathcal{P}_{\scriptsize{\mbox{obs}}}*
\mathcal{P}_{\scriptsize{\mbox{th}}}\right)(z) & = & 
\int_{-\infty}^{\infty}\mathcal{P}_{\scriptsize{\mbox{obs}}}(y)
\mathcal{P}_{\scriptsize{\mbox{th}}}(z\pm y)\,dy\,\nonumber
\\
& = &
\int_{-\infty}^{\infty}\mathcal{P}_{\scriptsize{\mbox{obs}}}(x\mp z)
\mathcal{P}_{\scriptsize{\mbox{th}}}(x)\, dx\,.
\label{pdf-conv}
\end{eqnarray}
where the plus (minus) sign refers to the difference (sum) of $x$ and $y$. 
Integrating this pdf from $(-\infty,0]$ we get our answer
\begin{equation}
P_\leq\equiv\int_{-\infty}^0\mathcal{P}(z)dz\,.
\label{Pconv>0}
\end{equation}
As a consistency check, notice that in the limit where observations are 
made with infinite precision, $\mathcal{P}_{\scriptsize{\mbox{obs}}}(y)$
becomes a delta function and we have:
\[
P_\leq=\int_{-\infty}^0\int_{-\infty}^{+\infty}\delta(y-x_0)
\mathcal{P}_{\scriptsize{\mbox{th}}}(z\pm y)dydz=
\int_{-\infty}^{\pm x_0}\mathcal{P}_{\scriptsize{\mbox{th}}}(x)dx
\]
which agrees with our previous definition. The reader must be careful,
though, not to think of Eq. (\ref{P>0}) as some lower bound to
Eq. (\ref{Pconv>0}). Since none of the pdf's appearing in
Eq. (\ref{pdf-conv}) are necessarily symmetric, a large distance from $x_0$
to the most probable (not the mean!) value $x$ would not, by itself,
constitute sufficient grounds to claim that the measured value of this
observable is ``unusual'' in any sense, simply because a large overlap
between the two pdf's can render the result usual according to
Eq. (\ref{Pconv>0}). 

To illustrate this point, let us calculate $P_\leq$ using Eq. (\ref{P>0})
and Eq. (\ref{Pconv>0}) for the pdf's which appear in Fig. 6. For
pedagogical reasons, we have chosen $\mathcal{P}_{\scriptsize{\mbox{th}}}$
and $\mathcal{P}_{\scriptsize{\mbox{obs}}}$ as positively (Maxwell
distribution) and negatively (Gumbel distribution) skewed pdf's,
respectively. The convolved distribution appears as the solid (black) line.
For these pdf's, Eq. (\ref{P>0}) gives $99.2\%$, while Eq. (\ref{Pconv>0})
gives $93.6\%$ of chance of $x$ being smaller than the observed $x_0$; all
pdf's were normalized to one. 
\begin{center}
\begin{figure}[H]
\begin{centering}
\includegraphics[scale=1]{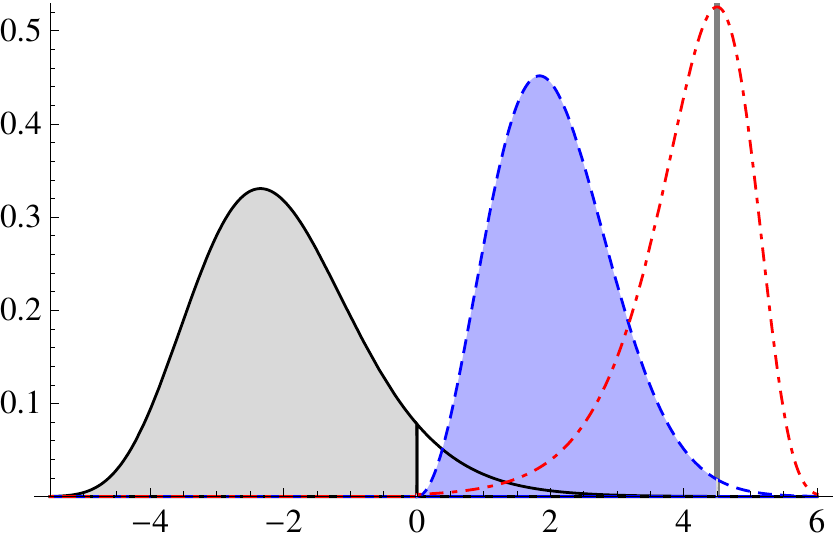}
\label{pdf-z}
\par\end{centering}
\caption{Probability density functions for $x$ (blue, dashed line), $y$ 
(red, dot-dashed line) and $z$ (solid line.) The shaded area gives the
probability of $x$ being smaller than $x_0$ (dashed vertical line.) See
the text for more details.}
\end{figure}
\par\end{center}

\section{A $\chi^2$ test of statistical isotropy}\label{sec:chi2-test}

Although the last example was constructed to emphasize an important 
feature of the formalism developed in \S\ref{sec:probabilities},
cosmological observables designed to measure deviations 
of either gaussianity or statistical isotropy will often follow asymmetric
distributions. The intrinsic uncertainties of cosmological measurements,
specially the ones originating from map cleaning procedures, may be crucial
when searching for any map's anomalies. We will now make a concrete
application of this formalism using the angular-planar chi-square test
developed in \S\ref{sec:angular-planar}

\subsection{$\Lambda$CDM model}

In the simplest realization of the $\Lambda$CDM model, the covariance 
matrix of temperature maps is determined by
$\langle a_{\ell_{1}m_{1}}a_{\ell_{2}m_{2}}\rangle=(-1)^{m_2}
C_{\ell_{1}}\delta_{\ell_{1}\ell_{2}}\delta_{m_{1},-m_{2}}$. Using this 
expression in (\ref{Clmell}), we find that
\begin{equation}
\mathcal{C}_{\ell}^{lm}\;\overset{\textrm{(SI)}}{=}\; C_{\ell}\delta_{l0}
\delta_{m0}\,.
\label{SI}
\end{equation}
On the other hand, if the only non-zero $\mathcal{C}_{\ell}^{lm}$'s are
given by $l=m=0$, then $\mathcal{C}_{\ell}^{00}=C_{\ell}$. Therefore, 
statistical isotropy is achieved if and only if the angular-planar power
spectrum is of the form (\ref{SI}). Since we are only interested in
nontrivial planar modulations, we will restrict our analysis to the 
cases where $l\neq0$, that is
\begin{equation}
\mathcal{C}_{\ell}^{lm}=0\,,\quad (l\geq2)\,
\label{nh}
\end{equation}
where, we remind the reader, the parity of $\ell$ comes from the symmetry (\ref{rr}).

For this particular model, we can also calculate the covariance matrix of 
the estimator (\ref{Clmell-hat}) explicitly. Using the null hypothesis
above, we find after some algebra that
\begin{equation}
\langle(\widehat{\mathcal{C}}_{\ell}^{\;lm})^{*}\widehat{\mathcal{C}}_{\ell}^{\;l'm'}
\rangle = 8\pi^{2}\sum_{\ell_{1},\ell_{2}}C_{\ell_{1}}C_{\ell_{2}}
\left(I_{\ell_{1}\ell_{2}}^{l,\ell}\right)^{2}\delta_{ll'}\delta_{mm'}\,.
\label{cov-matrix-Clmell}
\end{equation}
This matrix has some interesting properties. First, note that the planar
degrees of freedom are independent in this case (which justifies the
$(2l+1)^{-1}$) factor introduced in Eq. (\ref{chi2}).) Second, its
diagonal elements are given by the variance
$(\sigma^{lm}_\ell)^2=\langle(\widehat{\mathcal{C}}^{\;lm}_\ell)^*
\widehat{\mathcal{C}}^{\;lm}_\ell\rangle$, which now becomes
$m$-independent:
\begin{equation}
\left( \sigma^{lm}_{\ell} \right)^2 
\; \rightarrow \left( \sigma^{l}_\ell \right)^2=
8\pi^{2}\sum_{\ell_{1},\ell_{2}}C_{\ell_{1}}C_{\ell_{2}}
\left(I_{\ell_{1}\ell_{2}}^{l,\ell}\right)^{2}\,.
\label{Blell}
\end{equation}

Therefore, for the particular case of the $\Lambda$CDM model, the 
chi-square test (\ref{chi2}) gets even simpler. Using (\ref{nh}) and
(\ref{Blell}), we find
\begin{equation}
(\chi^2_\nu)^l_\ell=\frac{1}{2l+1}\sum_{m=-l}^l
\frac{|\widehat{\mathcal{C}}^{\;lm}_\ell|^2}{(\sigma^l_\ell)^2}\,.
\label{chi2-iso}
\end{equation}
It is now clear that if the data under analysis are really described by 
this model, then it must be true that
\[
\langle(\chi^2_\nu)^l_\ell\rangle=\frac{1}{2l+1}\sum_{m=-l}^l
\frac{\langle(\widehat{\mathcal{C}}^{\;lm}_\ell)^*
\widehat{\mathcal{C}}^{\;lm}_\ell\rangle}
{(\sigma^l_\ell)^2}=1\,
\]
where we have used (\ref{Blell}). This shows that any large deviation of 
this test from unity will be an indication of planar modulation in
temperature maps, up, of course, to error bars. For convenience, let us
define a new quantity as
\begin{equation}
\overline{\chi}^{\,l}_\ell\equiv(\chi^2_\nu)^l_\ell-1\,
\label{blell-barra}
\end{equation}
which will quantify anisotropies whenever $\overline{\chi}^{\,l}_\ell$ is 
significantly positive or negative.\\

This generalized chi-square test furnishes a complete prescription when
searching for planar modulations of temperature in CMB maps. We emphasize,
though, that for a given CMB map, the chi-square analysis must be done
entirely in terms of that map's data. Since we are performing a
model-independent test, we are not allowed to introduce fiducial biases in
the analysis (for example, by calculating $\sigma^l_{\ell}$ using
$C_\ell^{\scriptsize{\rm{model}}}$), which would only include our \textit{a
priori} prejudices about what the map's anisotropies should look like. 
Since the $C_\ell$'s are, by construction, a measure of statistical
isotropy. Consequently, an ``anomalous'' detection of $C_{\ell}$'s is by no
means a measure of statistical anisotropy, and it is this particular value
that should be used in (\ref{blell-barra}) if we want to find deviations of
isotropy, regardless of how high/low it is. 

\subsection{Searching for planar signatures in WMAP}

In order to apply the test (\ref{blell-barra}) to the 5-year WMAP data
\cite{Hinshaw:2008kr,Komatsu:2008hk}, we will define two new variables 
\[
x \equiv (\bar{\chi}^2)_{\ell(\scriptsize{\mbox{th}})}^{l} \, ,
\quad 
y \equiv (\bar{\chi}^2)_{\ell(\scriptsize{\mbox{obs}})}^{l} \, ,
\quad 
\]
which will be jointly analyzed using the formalism of section 
\S\ref{sec:probabilities}. Still, there remains the question of how to
obtain their pdf's. These functions can be obtained numerically provided
that the number of realizations of each variable is large enough, since 
in this case their histograms can be considered as piecewise constant
functions which approximate the real pdf's. For the case of the
(theoretical) variable $x$ defined above we have run $2\times10^{4}$ Monte
Carlo simulations of Gaussian and statistically isotropic CMB maps using
the $\Lambda\mbox{CDM}$ best-fit $C_{\ell}$'s provided by the WMAP team
\cite{lambda}.
With these maps we have then constructed $2\times10^{4}$ realizations of
the variable $x$. 

The simulation of the (observational) variable $y$ is more difficult, and 
depends on the way we estimate contamination from residual foregrounds in
CMB maps. As is well-known, not only instrumental noise, but systematic
errors (e.g., in the map-making process), the inhomogeneous scanning of the
sky ({\it i.e.}, the exposure function of the probe), or unremoved
foreground emissions (even after applying a cut-sky mask) could corrupt --
at distinct levels -- the CMB data.

Foreground contamination, on the other hand, may have several different 
sources, many of which are far beyond our present scopes. However, since
different teams apply distinct procedures on the raw data in order to
produce a final map, we will make the hypothesis that maps cleaned by
different teams represent -- to a good extent -- ``independent'' CMB maps.
Therefore, we can estimate the residual foreground contaminations by
comparing these different foreground-cleaned maps.
\begin{table}[h]
\begin{centering}
\begin{tabular}{cc}
\toprule 
Full sky maps & References\tabularnewline
\midrule
\midrule 
Hinshaw \textit{et. al.} &
\cite{Hinshaw:2008kr,Hinshaw:2006ia}\tabularnewline
\midrule 
de Oliveira-Costa \textit{et. al.} & \cite{deOliveiraCosta:2006zj}
\tabularnewline
\midrule 
Kim \textit{et. al.} & \cite{Kim:2008zh}\tabularnewline
\midrule 
Park \textit{et. al.}  & \cite{Park:2006dv}\tabularnewline
\midrule 
Delabrouille \textit{et. al.} & \cite{Delabrouille:2008qd}\tabularnewline
\bottomrule
\end{tabular}

\caption{Full-sky foreground cleaned CMB maps from WMAP data used in our 
analysis to estimate the variable $y$ (see the text for more details.) 
Note that the reference \cite{Kim:2008zh} includes the analysis 
of maps from the three and five years WMAP releases.}
\label{tab:mapas-ceu-int}
\par\end{centering}
\end{table}

In fact, the WMAP science team has made substantial efforts to improve the 
data products by minimizing the contaminating effects caused by diffuse
galactic foregrounds, astrophysical point-sources, artifacts from the
instruments and measurement process, and systematic
errors~\cite{Jarosik:2006ib,Gold:2008kp}. As a result, multi-frequency
foreground-cleaned full-sky CMB maps were produced, named Internal Linear
Combination maps, corresponding to three and five year WMAP
data~\cite{Hinshaw:2008kr,Hinshaw:2006ia}. To account for the mentioned
randomness, systematic, and contaminating effects of the CMB data, we will
use in our analyses several full-sky foreground-cleaned CMB maps, listed in
Table~\ref{tab:mapas-ceu-int}, which were produced using both the three and
five year WMAP data.\\

The prescription we adopt to determine the distribution of the 
observational variable $y$ is as follows: we simulate Gaussian random
$a_{\ell m}$'s in such a way that {\it their central values are given by
the five year ILC5 data} \cite{Hinshaw:2006ia,Hinshaw:2008kr},
and with a variance which is estimated from the \textit{sample standard
deviation} of all the maps listed in Table~\ref{tab:mapas-ceu-int}. 
So, for example, suppose we have $n$ different full-sky temperature maps 
at hand and we want to estimate the randomness inherent in the
determination of, let's say, $a_{32}$. Therefore, we take
\begin{equation}
\mathcal{N}(a_{32}^{\scriptsize{\mbox{ILC5}}},\sigma_{32})\;\rightarrow\; 
a_{32} \, ,
\label{protocolo-1}
\end{equation}
with
\begin{equation}
\sigma_{32}=\sqrt{\frac{1}{n-1}\sum_{i=1}^{n}(a_{32}^{i}-\bar{a}_{32})^{2}}
\qquad\mbox{and}
\qquad\bar{a}_{32}=\frac{1}{n}\sum_{i=1}^{n}a_{32}^{i} \, ,
\label{protocolo-2}
\end{equation}
where $\mathcal{N}(\mu,\sigma)$ represents a Gaussian distribution
with mean $\mu$ and standard deviation $\sigma$. Note that if the residual
contamination is indeed weak, then the sample variance above will be small,
and our procedure will reduce to the standard way of calculating
probabilities. As for the use of a Gaussian in (\ref{protocolo-1}), this choice was
dictated not only by simplicity, but rather by the fact that the propagation of uncertainties 
in physical experiments are usually assumed to follow a normal distribution. Note however
that there are some instrumental uncertainties, such as beam or gain uncertainties, which will 
not in general follow normal distributions. In fact, some of them may even fail to be additive,
meaning that our convolution formula will be inapplicable in these cases. In our analysis, we 
have focused only on foreground residuals, where the normality hypothesis is reasonable\footnote{We 
thank the referee for pointing out this important aspect of the formalism.}.

\subsection{Full-sky maps}
\begin{figure}[H]
\begin{center}
\includegraphics[scale=0.54]{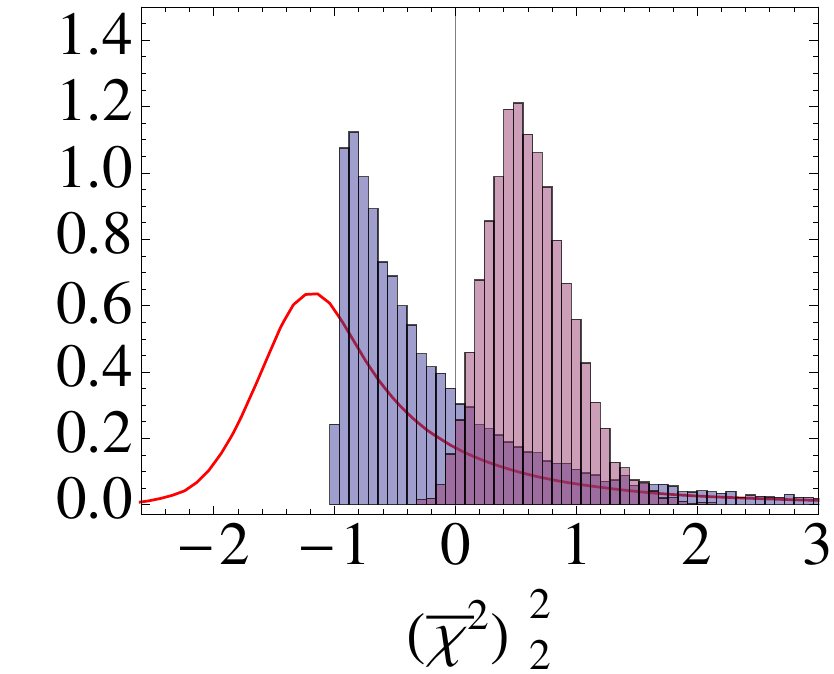} 
\includegraphics[scale=0.54]{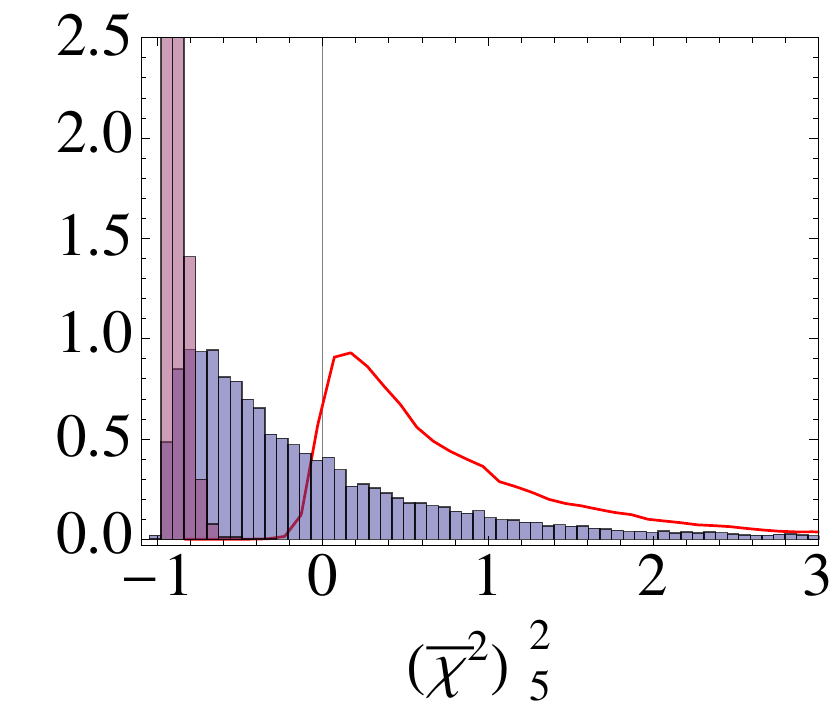}
\includegraphics[scale=0.54]{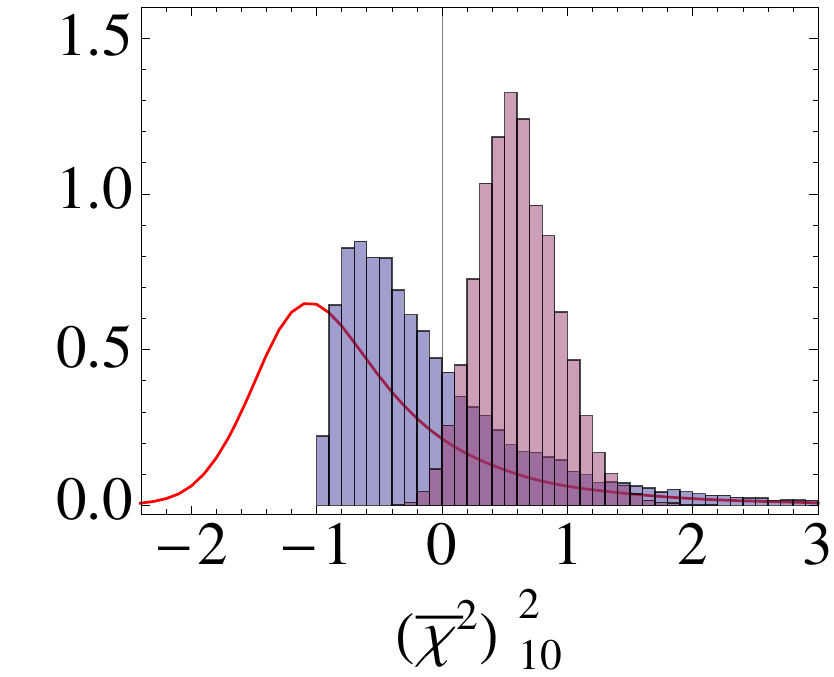}

\includegraphics[scale=0.54]{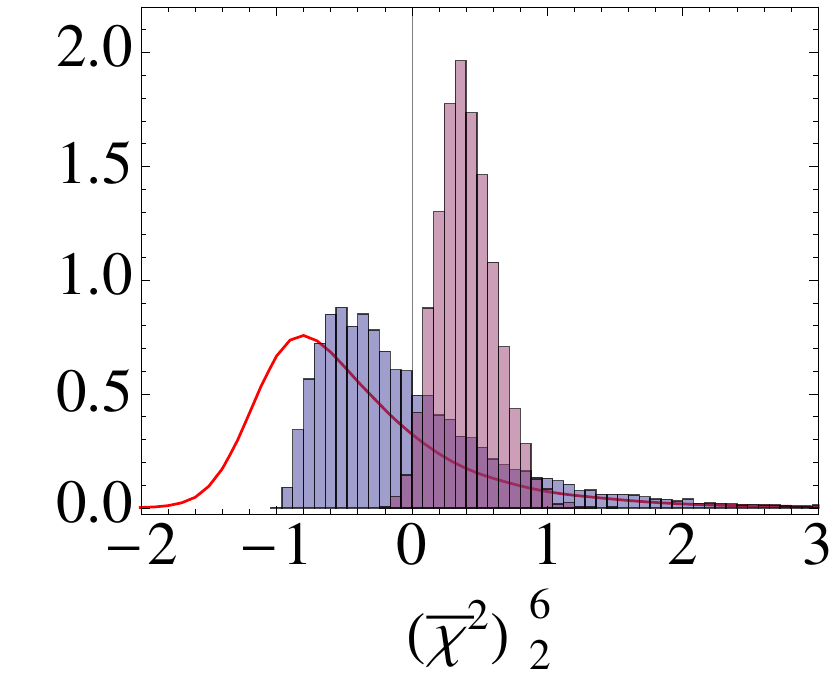}
\includegraphics[scale=0.54]{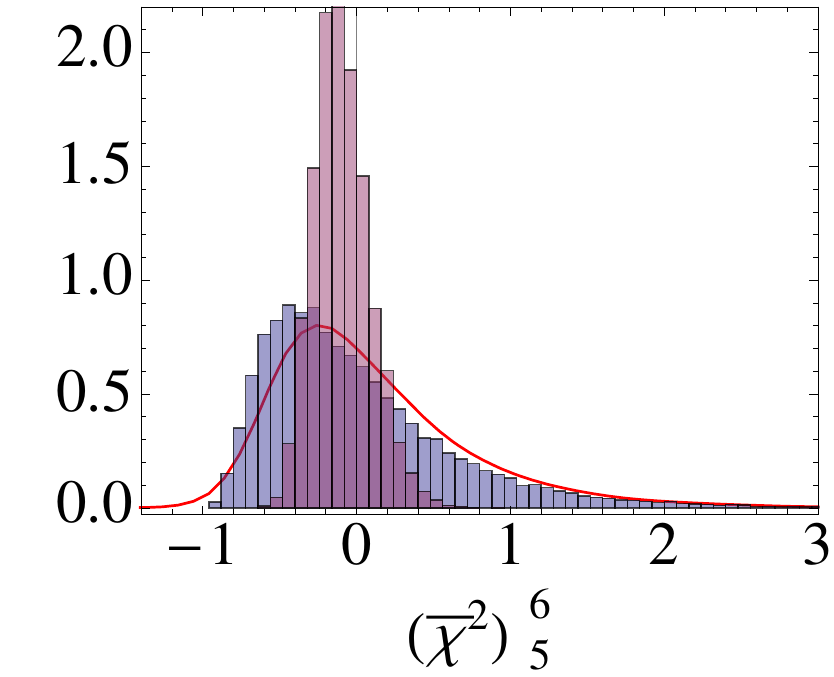}
\includegraphics[scale=0.54]{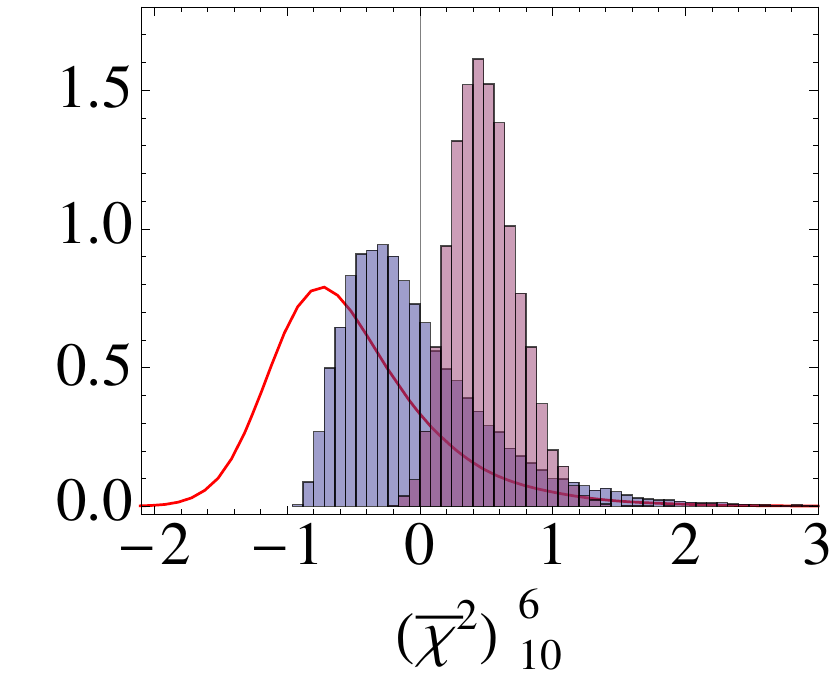}

\includegraphics[scale=0.54]{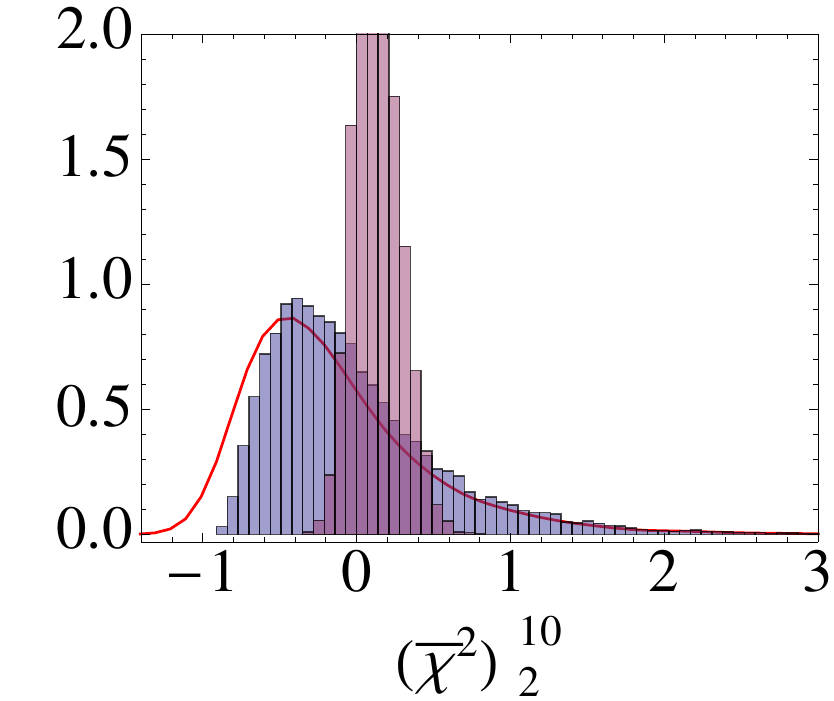}
\includegraphics[scale=0.54]{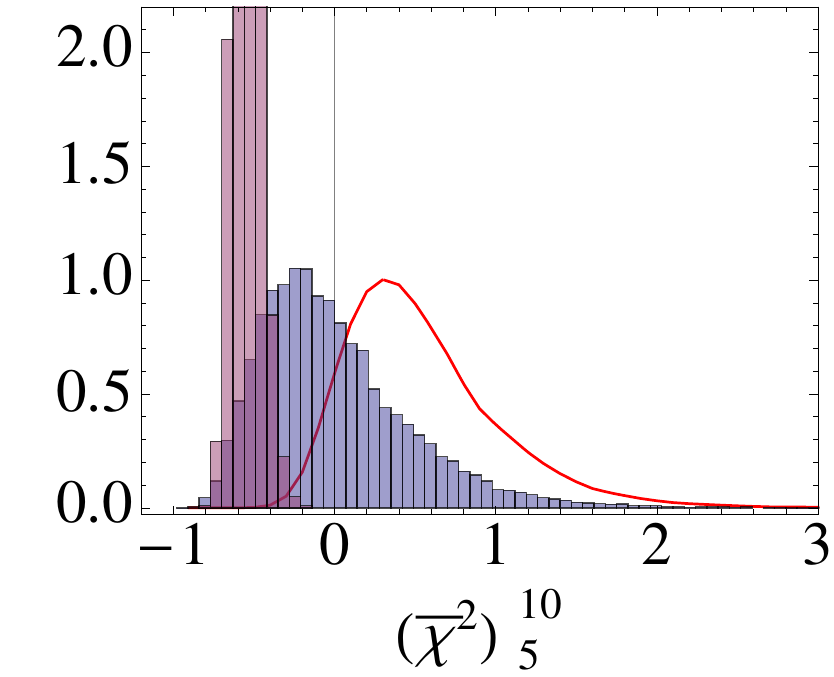}
\includegraphics[scale=0.54]{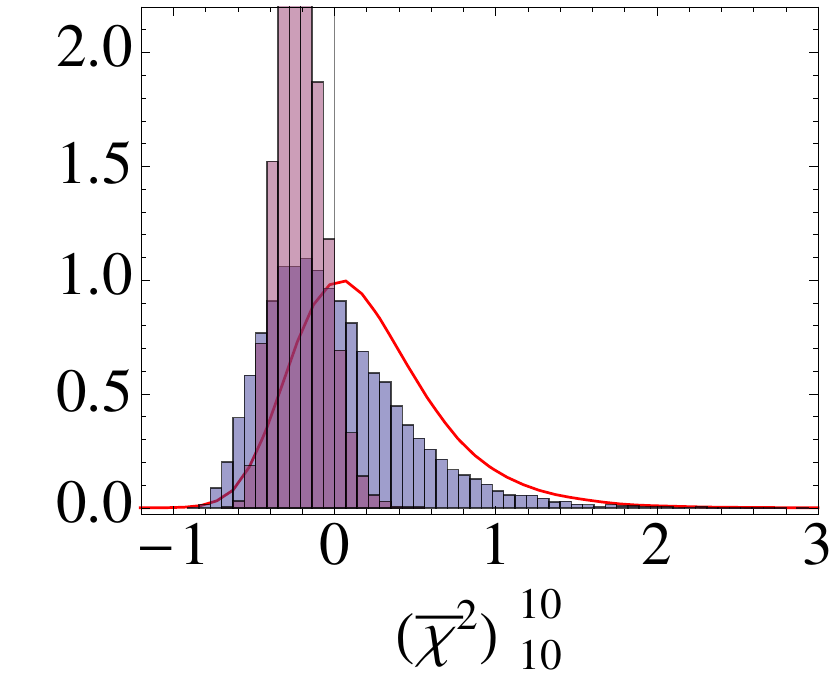}

\caption{Histograms for
$(\bar{\chi}^2)_{\ell(\scriptsize{\mbox{th}})}^{l}$
(blue), $(\bar{\chi}^2)_{\ell(\scriptsize{\mbox{obs}})}^{l}$
(purple) and for the difference
$(\bar{\chi}^2)_{\ell(\scriptsize{\mbox{th}})}^{l} -
(\bar{\chi}^2)_{\ell(\scriptsize{\mbox{obs}})}^{l}$
(solid, red line). We show only a few representative figures. The final
probabilities are shown in Table \ref{tab:prob-finais-ceuint} and
correspond to the area under the solid curve from $-\infty$ to $0$. All
pdf's are normalized to 1.
\label{fig:hist-ceuint}}
\end{center}
\end{figure}

Following this procedure, we have used the full-sky maps shown in Table
(\ref{tab:mapas-ceu-int}) to construct $10^{4}$ Gaussian random $a_{\ell
m}$'s, which were then used to calculate $10^{4}$ realizations of
$y=(\bar{\chi}^2)^l_{\ell({\rm obs})}$. With those variables we constructed
histograms which, together with the histograms for the (full-sky) variable
$x$, were used to calculate the final probability (\ref{Pconv>0}). We have
restricted our analysis to the range of values $(\ell,l)\in[2,10]$, since
the low multipolar sector ({\it i.e.}, large angular scales) is where most
of the anomalies were reported. The resulting histograms and pdf's are
shown in Fig. \ref{fig:hist-ceuint}, and the final probabilities we
obtained are show in Table \ref{tab:prob-finais-ceuint}.\\

\begin{table}
\begin{center}
\begin{longtable}{cccccccccc}
\toprule 
$l\backslash\ell$ & \multicolumn{1}{c}{2} & \multicolumn{1}{c}{3} & 4 & 5 &
6 & 7 & 8 & 9 & 10\tabularnewline
\midrule
\midrule 
2 & 81.1\% & 73.7\% & 54.1\% & \textbf{6.1}\% & 80.7\% & 46.4\% & 36.9\% &
47.5\% & 81.8\% \tabularnewline
\midrule 
4 & 74.0\% & 72.6\% & 55.0\% & 39.6\% & 74.2\% & \textbf{93.2\%} & 51.6\% &
55.4\% & 56.3\%\tabularnewline
\midrule 
6 & 78.1\% & 80.7\% & 69.3\% & 52.3\% & 33.6\% & 80.0\% & \textbf{95.0}\% &
50.3\% & 82.2\%\tabularnewline
\midrule 
8 & 63.5\% & 87.7\% & 18.8\% & 51.5\% & 21.4\% & 66.4\% & 31.6\% & 27.5\% &
82.3\%\tabularnewline
\midrule 
10 & 67.9\% & 50.0\% & 61.0\% & \textbf{8.7}\% & 37.7\% & 59.5\% & 36.6\% &
29.2\% & 35.7\%\tabularnewline
\bottomrule
\end{longtable}
\caption{Final probabilities of obtaining, in a random $\Lambda$CDM
Universe, a chi-square value smaller or equal to
$(\bar{\chi}^2)_{\ell(\scriptsize{\rm obs})}^{l}$, as given by 
full-sky temperature maps.
\label{tab:prob-finais-ceuint}}
\end{center}
\end{table}


Overall, our results show no significant planar deviations of anisotropy in
WMAP data. The most unlikely individual values in Table
(\ref{tab:prob-finais-ceuint}) are in the sectors $(l,\ell)$ given by
$(2,5)$, $(10,5)$, $(4,7)$ and $(6,8)$, and are all above a relative chance
of 5\% of either being too negative [$\left(2,5\right)$,
$\left(10,5\right)$] or too positive [$(4,7)$, $(6,8)$]. 
However, it is perhaps worth mentioning that not only the individual
values of $(\bar{\chi}^2)^l_\ell$ are relevant, their coherence over a
range of angular or planar momenta also carries interesting information.
So, for example, a set of $(\bar{\chi}^2)^l_\ell$'s which are all
individually within the cosmic variance bounds, but which are all positive
(or negative) can be an indication of an excess (or lack) of planar
modulation. This type of coherent behavior appears in the following cases:
$(\bar{\chi}^2)^l_2$, $(\bar{\chi}^2)^l_3$ and, to a lesser extent, 
$(\bar{\chi}^2)^4_\ell$ -- see Table \ref{tab:prob-finais-ceuint}. The
angular quadrupole $\ell=2$, as well as the angular octupole $\ell=3$, have
all positive planar spectra (for all values of $l$ which we were able to
compute), indicated by probabilities larger than 50\%. The planar
hexadecupole $l=4$ also has 8 out of 9 angular spectra assuming positive
values (only $\ell=5$ is negative). 

The data analyzed in this Section relates to the full-sky maps, which are
certainly still affected by residual galactic foregrounds. The reader
interested in the complete analysis, including data from masked CMB maps,
can check reference \cite{Abramo:2009fe}.


\newpage
\section{Conclusions}\label{sec:conclusions}

We know our Universe is not perfectly Gaussian, or homogeneous, or isotropic. 
The deviations from an idealized picture (or the lack thereof), whether 
predictably small or surprisingly large, can tell us a great deal about the 
Universe we live in. Since the types of physical mechanisms behind deviations 
from perfect gaussianity, homogeneity or isotropy are typically very different, 
we should try to measure these individual features separately --
whenever possible or practical.

Recently it has been suggested that some of the most discussed anomalies
in the CMB can be explained away \cite{Francis:2009pt},
or that the evidence for them is statistically weak \cite{Bennett:2010jb}.
But even if it turns out that our Universe is a plain vanilla kind of place, where 
everything goes according to the inflationary theorist's dreams, we would still need
to analyze it with tools that allow us to check the standard picture against the data.
In addition, local physics (related to the solar system, or our galaxy), as well
as instrumental quirks, tend to leave imprints on the CMB which are
clearly anisotropic, but have a certain coherence which can be detected,
and possibly corrected for, with the help of these checks.

However, in an era where at least the large-scale maps of the CMB are likely to 
remain basically unchanged, we should be careful not to over analize the data 
with the benefit of an ever greater hindsight (put another way, {\it a posteriori}
conundrums only get worse with time.) This can only be achieved if we find 
natural and generally agreed-upon classifications of the types of deviations that 
may occur, without too much guidance from what the data is telling us. 
We believe that focusing on the possible underlying symmetries, with 
perhaps some guidance from group-theoretic arguments, is one
way to settle these issues. We have presented a few methods along
these lines, one using multipole vectors, the other using a natural 
generalization of the two-point correlation function -- and other methods
have been presented in this Review.

Perhaps the best indication that we are on the right track is the fact that
most of these methods are applicable in other areas of physics and
astronomy -- and that in some cases we have adapted tests of anisotropy
from other areas, such as scattering theory and the theory of angular
momentum in quantum mechanics. So, even if these anomalies eventually
perish, they will be survived by the powerful methods that have been
devised to test them.

\subsection*{Acknowledgements}

We would like to thank Glenn Starkman and Yuri Shtanov for enlightening
discussions during the development of this work. This work was supported by
FAPESP and CNPq. 

\bibliographystyle{h-physrev}
\addcontentsline{toc}{section}{\refname}\bibliography{reviewaa}

\begin{thebibliography}{10}

\bibitem{Mukhanov:05}
V.~Mukhanov,
\newblock {\em Physical Foundations of Cosmology} (Cambridge University Press,
  2005).

\bibitem{Weinberg:2008zzc}
S.~Weinberg,
\newblock {\em Cosmology} (Oxford Univ. Press, 2008).

\bibitem{Peter:2009zzc}
P.~Peter and J.-P. Uzan,
\newblock {\em Primordial Cosmology} (Oxford Univ. Press, 2009).

\bibitem{Hu:1997mn}
W.~Hu, U.~Seljak, M.~J. White, and M.~Zaldarriaga,
\newblock Phys. Rev. {\bf D57}, 3290 (1998), astro-ph/9709066.

\bibitem{Hu:2002aa}
W.~Hu,
\newblock Ann. Phys. {\bf 303}, 203 (2003), astro-ph/0210696.

\bibitem{Dodelson:03}
S.~Dodelson,
\newblock {\em Modern Cosmology} (Academic Press, 2003).

\bibitem{Maldacena:2002vr}
J.~M. Maldacena,
\newblock JHEP {\bf 05}, 013 (2003), astro-ph/0210603.

\bibitem{Weinberg:2005vy}
S.~Weinberg,
\newblock Phys. Rev. {\bf D72}, 043514 (2005), hep-th/0506236.

\bibitem{Weinberg:2006ac}
S.~Weinberg,
\newblock Phys. Rev. {\bf D74}, 023508 (2006), hep-th/0605244.

\bibitem{Huffenberger:2004gm}
K.~M. Huffenberger and U.~Seljak,
\newblock New Astron. {\bf 10}, 491 (2005), astro-ph/0408066.

\bibitem{Komatsu:2008hk}
WMAP, E.~Komatsu {\em et~al.},
\newblock Astrophys. J. Suppl. {\bf 180}, 330 (2009), 0803.0547.

\bibitem{Bernui:2008ei}
A.~Bernui and M.~J. Reboucas,
\newblock Phys. Rev. {\bf D79}, 063528 (2009), 0806.3758.

\bibitem{Mukhanov:1990me}
V.~F. Mukhanov, H.~A. Feldman, and R.~H. Brandenberger,
\newblock Phys. Rept. {\bf 215}, 203 (1992).

\bibitem{Hinshaw:2008kr}
WMAP, G.~Hinshaw {\em et~al.},
\newblock Astrophys. J. Suppl. {\bf 180}, 225 (2009), 0803.0732.

\bibitem{Tegmark:2003uf}
SDSS, M.~Tegmark {\em et~al.},
\newblock Astrophys. J. {\bf 606}, 702 (2004), astro-ph/0310725.

\bibitem{Cole:2005sx}
The 2dFGRS, S.~Cole {\em et~al.},
\newblock Mon. Not. Roy. Astron. Soc. {\bf 362}, 505 (2005), astro-ph/0501174.

\bibitem{Perlmutter:1998np}
Supernova Cosmology Project, S.~Perlmutter {\em et~al.},
\newblock Astrophys. J. {\bf 517}, 565 (1999), astro-ph/9812133.

\bibitem{Riess:1998cb}
Supernova Search Team, A.~G. Riess {\em et~al.},
\newblock Astron. J. {\bf 116}, 1009 (1998), astro-ph/9805201.

\bibitem{Wood-Vasey:2007jb}
ESSENCE, W.~M. Wood-Vasey {\em et~al.},
\newblock Astrophys. J. {\bf 666}, 694 (2007), astro-ph/0701041.

\bibitem{Bennett:2003bz}
WMAP, C.~L. Bennett {\em et~al.},
\newblock Astrophys. J. Suppl. {\bf 148}, 1 (2003), astro-ph/0302207.

\bibitem{Smoot:1992td}
G.~F. Smoot {\em et~al.},
\newblock Astrophys. J. {\bf 396}, L1 (1992).

\bibitem{deOliveiraCosta:2003pu}
A.~{de Oliveira-Costa}, M.~Tegmark, M.~Zaldarriaga, and A.~Hamilton,
\newblock Phys. Rev. {\bf D69}, 063516 (2004), astro-ph/0307282.

\bibitem{Copi:2003kt}
C.~J. Copi, D.~Huterer, and G.~D. Starkman,
\newblock Phys. Rev. {\bf D70}, 043515 (2004), astro-ph/0310511.

\bibitem{Schwarz:2004gk}
D.~J. Schwarz, G.~D. Starkman, D.~Huterer, and C.~J. Copi,
\newblock Phys. Rev. Lett. {\bf 93}, 221301 (2004), astro-ph/0403353.

\bibitem{Land:2005ad}
K.~Land and J.~Magueijo,
\newblock Phys. Rev. Lett. {\bf 95}, 071301 (2005), astro-ph/0502237.

\bibitem{Copi:2005ff}
C.~J. Copi, D.~Huterer, D.~J. Schwarz, and G.~D. Starkman,
\newblock Mon. Not. Roy. Astron. Soc. {\bf 367}, 79 (2006), astro-ph/0508047.

\bibitem{Tegmark:2003ve}
M.~Tegmark, A.~{de Oliveira-Costa}, and A.~Hamilton,
\newblock Phys. Rev. {\bf D68}, 123523 (2003), astro-ph/0302496.

\bibitem{Abramo:2009fe}
L.~R. Abramo, A.~Bernui, and T.~S. Pereira,
\newblock JCAP {\bf 0912}, 013 (2009), 0909.5395.

\bibitem{Bernui:2005pz}
A.~Bernui, B.~Mota, M.~J. Reboucas, and R.~Tavakol,
\newblock Astron. Astrophys. {\bf 464}, 479 (2007), astro-ph/0511666.

\bibitem{Bernui:2008cr}
A.~Bernui,
\newblock Phys. Rev. {\bf D78}, 063531 (2008), 0809.0934.

\bibitem{Eriksen:2003db}
H.~K. Eriksen, F.~K. Hansen, A.~J. Banday, K.~M. Gorski, and P.~B. Lilje,
\newblock Astrophys. J. {\bf 605}, 14 (2004), astro-ph/0307507.

\bibitem{Eriksen:2007pc}
H.~K. Eriksen, A.~J. Banday, K.~M. Gorski, F.~K. Hansen, and P.~B. Lilje,
\newblock Astrophys. J. {\bf 660}, L81 (2007), astro-ph/0701089.

\bibitem{Pietrobon:2009qg}
D.~Pietrobon {\em et~al.},
\newblock (2009), arXiv:0905.3702.

\bibitem{Abramo:2006hs}
L.~R. Abramo, L.~S. Jr., and C.~A. Wuensche,
\newblock Phys. Rev. {\bf D74}, 083515 (2006), astro-ph/0605269.

\bibitem{Abramo:2006gw}
L.~R. Abramo, A.~Bernui, I.~S. Ferreira, T.~Villela, and C.~A. Wuensche,
\newblock Phys. Rev. {\bf D74}, 063506 (2006), astro-ph/0604346.

\bibitem{Hanson:2009gu}
D.~Hanson and A.~Lewis,
\newblock Phys. Rev. {\bf D80}, 063004 (2009), 0908.0963.

\bibitem{Pullen:2007tu}
A.~R. Pullen and M.~Kamionkowski,
\newblock Phys. Rev. {\bf D76}, 103529 (2007), 0709.1144.

\bibitem{Pereira:2007yy}
T.~S. Pereira, C.~Pitrou, and J.-P. Uzan,
\newblock JCAP {\bf 0709}, 006 (2007), 0707.0736.

\bibitem{Pitrou:2008gk}
C.~Pitrou, T.~S. Pereira, and J.-P. Uzan,
\newblock JCAP {\bf 0804}, 004 (2008), 0801.3596.

\bibitem{Gumrukcuoglu:2007bx}
A.~E. Gumrukcuoglu, C.~R. Contaldi, and M.~Peloso,
\newblock JCAP {\bf 0711}, 005 (2007), 0707.4179.

\bibitem{Varshalovich:1988ye}
D.~A. Varshalovich, A.~N. Moskalev, and V.~K. Khersonsky,
\newblock SINGAPORE, SINGAPORE: WORLD SCIENTIFIC (1988) 514p.

\bibitem{Uzan:2008qp}
J.-P. Uzan, C.~Clarkson, and G.~F.~R. Ellis,
\newblock Phys. Rev. Lett. {\bf 100}, 191303 (2008), 0801.0068.

\bibitem{Prunet:2004zy}
S.~Prunet, J.-P. Uzan, F.~Bernardeau, and T.~Brunier,
\newblock Phys. Rev. {\bf D71}, 083508 (2005), astro-ph/0406364.

\bibitem{Ferreira:1997wd}
P.~G. Ferreira and J.~Magueijo,
\newblock Phys. Rev. {\bf D56}, 4578 (1997), astro-ph/9704052.

\bibitem{Fry:1985ri}
J.~N. Fry and A.~L. Melott,
\newblock NSF-ITP-84-143.

\bibitem{Luo:1993xx}
X.~chun Luo,
\newblock Astrophys. J. {\bf 427}, L 71 (1994), astro-ph/9312004.

\bibitem{Spergel:1999xn}
D.~N. Spergel and D.~M. Goldberg,
\newblock Phys. Rev. {\bf D59}, 103001 (1999), astro-ph/9811252.

\bibitem{Hu:2001fa}
W.~Hu,
\newblock Phys. Rev. {\bf D64}, 083005 (2001), astro-ph/0105117.

\bibitem{Hivon:2001jp}
E.~Hivon {\em et~al.},
\newblock (2001), astro-ph/0105302.

\bibitem{Ackerman:2007nb}
L.~Ackerman, S.~M. Carroll, and M.~B. Wise,
\newblock Phys. Rev. {\bf D75}, 083502 (2007), astro-ph/0701357.

\bibitem{Shtanov:2009wp}
Y.~Shtanov and H.~Pyatkovska,
\newblock Phys. Rev. {\bf D80}, 023521 (2009), 0904.1887.

\bibitem{Luminet:2003dx}
J.~P. Luminet, J.~Weeks, A.~Riazuelo, R.~Lehoucq, and J.~P. Uzan,
\newblock Nature. {\bf 425}, 593 (2003), astro-ph/0310253.

\bibitem{Riazuelo:2003ud}
A.~Riazuelo, J.~Weeks, J.-P. Uzan, R.~Lehoucq, and J.-P. Luminet,
\newblock Phys. Rev. {\bf D69}, 103518 (2004), astro-ph/0311314.

\bibitem{HipolitoRicaldi:2005eh}
W.~S. Hipolito-Ricaldi and G.~I. Gomero,
\newblock Phys. Rev. {\bf D72}, 103008 (2005), astro-ph/0507238.

\bibitem{Kahniashvili:2008sh}
T.~Kahniashvili, G.~Lavrelashvili, and B.~Ratra,
\newblock Phys. Rev. {\bf D78}, 063012 (2008), 0807.4239.

\bibitem{Kahniashvili:2008hx}
T.~Kahniashvili, Y.~Maravin, and A.~Kosowsky,
\newblock Phys. Rev. {\bf D80}, 023009 (2009), 0806.1876.

\bibitem{Bernui:2008ve}
A.~Bernui and W.~S. Hipolito-Ricaldi,
\newblock Mon. Not. Roy. Astron. Soc. {\bf 389}, 1453 (2008), 0807.1076.

\bibitem{Caprini:2009vk}
C.~Caprini, F.~Finelli, D.~Paoletti, and A.~Riotto,
\newblock JCAP {\bf 0906}, 021 (2009), 0903.1420.

\bibitem{Seshadri:2009sy}
T.~R. Seshadri and K.~Subramanian,
\newblock Phys. Rev. Lett. {\bf 103}, 081303 (2009), 0902.4066.

\bibitem{Gordon:2005ai}
C.~Gordon, W.~Hu, D.~Huterer, and T.~M. Crawford,
\newblock Phys. Rev. {\bf D72}, 103002 (2005), astro-ph/0509301.

\bibitem{Campanelli:2007qn}
L.~Campanelli, P.~Cea, and L.~Tedesco,
\newblock Phys. Rev. {\bf D76}, 063007 (2007), 0706.3802.

\bibitem{Campanelli:2009tk}
L.~Campanelli,
\newblock (2009), arXiv:0907.3703.

\bibitem{Erickcek:2008sm}
A.~L. Erickcek, M.~Kamionkowski, and S.~M. Carroll,
\newblock Phys. Rev. {\bf D78}, 123520 (2008), 0806.0377.

\bibitem{Erickcek:2009at}
A.~L. Erickcek, C.~M. Hirata, and M.~Kamionkowski,
\newblock (2009), arXiv:0907.0705.

\bibitem{Donoghue:2007ze}
J.~F. Donoghue, K.~Dutta, and A.~Ross,
\newblock Phys. Rev. {\bf D80}, 023526 (2009), astro-ph/0703455.

\bibitem{Kawasaki:2008sn}
M.~Kawasaki, K.~Nakayama, T.~Sekiguchi, T.~Suyama, and F.~Takahashi,
\newblock JCAP {\bf 0811}, 019 (2008), 0808.0009.

\bibitem{Kawasaki:2008pa}
M.~Kawasaki, K.~Nakayama, T.~Sekiguchi, T.~Suyama, and F.~Takahashi,
\newblock JCAP {\bf 0901}, 042 (2009), 0810.0208.

\bibitem{Hajian:2003qq}
A.~Hajian and T.~Souradeep,
\newblock Astrophys. J. {\bf 597}, L5 (2003), astro-ph/0308001.

\bibitem{Hajian:2004zn}
A.~Hajian, T.~Souradeep, and N.~J. Cornish,
\newblock Astrophys. J. {\bf 618}, L63 (2004), astro-ph/0406354.

\bibitem{Hajian:2005jh}
A.~Hajian and T.~Souradeep,
\newblock (2005), astro-ph/0501001.

\bibitem{Pereira:2009kg}
T.~S. Pereira and L.~R. Abramo,
\newblock Phys. Rev. {\bf D80}, 063525 (2009), 0907.2340.

\bibitem{Weeks:2004cz}
J.~R. Weeks,
\newblock (2004), astro-ph/0412231.

\bibitem{Katz:2004nj}
G.~Katz and J.~Weeks,
\newblock Phys. Rev. {\bf D70}, 063527 (2004), astro-ph/0405631.

\bibitem{EHobson}
E.~Hobson,
\newblock {\em The Theory of Spherical and Ellipsoidal Harmonics.} (Chelsea
  Pub. Co., 1955).

\bibitem{Dennis:2007jk}
M.~R. Dennis and K.~Land,
\newblock (2007), arXiv:0704.3657.

\bibitem{Yuri}
See \cite{Pereira:2009kg} for more details. Incidentally, the functional
  dependence of Eq.(\ref{fc-aniso}) differs from that originaly shwon in
  \cite{Pereira:2009kg}, which does not correctly take the planar dependence
  into account. We thank Yuri Shtanov for pointing this out.

\bibitem{Ash}
R.~B. Ash,
\newblock {\em Basic Probability Theory.} (Dover, 2008).

\bibitem{lambda}
http://lambda.gsfc.nasa.gov/.

\bibitem{Hinshaw:2006ia}
WMAP, G.~Hinshaw {\em et~al.},
\newblock Astrophys. J. Suppl. {\bf 170}, 288 (2007), astro-ph/0603451.

\bibitem{deOliveiraCosta:2006zj}
A.~{de Oliveira-Costa} and M.~Tegmark,
\newblock Phys. Rev. {\bf D74}, 023005 (2006), astro-ph/0603369.

\bibitem{Kim:2008zh}
J.~Kim, P.~Naselsky, and P.~R. Christensen,
\newblock Phys. Rev. {\bf D77}, 103002 (2008), 0803.1394.

\bibitem{Park:2006dv}
C.-G. Park, C.~Park, and J.~R.~I. Gott,
\newblock Astrophys. J. {\bf 660}, 959 (2007), astro-ph/0608129.

\bibitem{Delabrouille:2008qd}
J.~Delabrouille {\em et~al.},
\newblock (2008), arXiv:0807.0773.

\bibitem{Jarosik:2006ib}
WMAP, N.~Jarosik {\em et~al.},
\newblock Astrophys. J. Suppl. {\bf 170}, 263 (2007), astro-ph/0603452.

\bibitem{Gold:2008kp}
WMAP, B.~Gold {\em et~al.},
\newblock Astrophys. J. Suppl. {\bf 180}, 265 (2009), 0803.0715.

\bibitem{Francis:2009pt}
C.~L. Francis and J.~A. Peacock,
\newblock (2009), arXiv:0909.2495.

\bibitem{Bennett:2010jb}
C.~L. Bennett {\em et~al.},
\newblock (2010), arXiv:1001.4758.

\end{thebibliography}


\appendix

\section{Geometrical identities and derivations}

\subsection*{Wigner $D$-functions}
From the unitarity of the rotation operators $D(R)$, we have
\[
\sum_{m}D_{m'm}^{\ell}(\omega_{1})D_{m''m}^{\ell}(\omega_{2})=
D_{m'm''}^{\ell}(\omega_{1}\omega_{2})\,.
\]
If $\omega_{2}=\omega_{1}^{-1}$, then using the identity $D_{m''m}^{\ell}(\omega^{-1})=D_{m''m}^{*\ell}(\omega)$,
we find
\[
\sum_{m}D_{m'm}^{\ell}(\omega_{1})D_{m''m}^{*\ell}(\omega_{1})
=\delta_{m'm''}\,.
\]

\subsection*{Gaunt integral}
The definition of the Gaunt integral used in this paper is
\[
\mathcal{G}_{m_{1}m_{2}m_{3}}^{\ell_{1}\ell_{2}\ell_{3}}=(-1)^{m_{1}}\int
d^2{\hat{n}}\;Y_{\ell_{1},-m_{1}}(\hat{n})Y_{\ell_{2}m_{2}}(\hat{n})
Y_{\ell_{2}m_{2}}(\hat{n})\,.
\]

\subsection*{3-j symbols}
We present here some useful identities related to the 3-j symbols: 
\begin{itemize}
\item Isotropic limit 
\[
\left(\begin{array}{ccc}
l_{1} & l_{2} & 0\\
m_{1} & m_{2} &
0\end{array}\right)=\frac{(-1)^{l_{2}-m_{1}}}{\sqrt{2l_{1}+1}}\delta_{l_{1}
l_{2}}\delta_{m_{1},-m_{2}}\,.
\]
\item Parity and permutations 
\begin{eqnarray*}
\left(\begin{array}{ccc}
l_{1} & l_{2} & l\\
m_{1} & m_{2} & m\end{array}\right) & = & \left(\begin{array}{ccc}
l & l_{1} & l_{2}\\
m & m_{1} & m_{2}\end{array}\right)\\
 & = & (-1)^{l_{1}+l_{2}+l}\left(\begin{array}{ccc}
l_{2} & l_{1} & l\\
m_{2} & m_{1} & m\end{array}\right)\\
 & = & (-1)^{l_{1}+l_{2}+l}\left(\begin{array}{ccc}
l_{1} & l_{2} & l\\
-m_{1} & -m_{2} & -m\end{array}\right)\,.
\end{eqnarray*}
\item Orthogonality 
\begin{eqnarray*}
\sum_{m_{1}=-l_{1}}^{l_{1}}\sum_{m_{2}=-l_{2}}^{l_{2}}\left(\begin{array}{
ccc}
l_{1} & l_{2} & l_{3}\\
m_{1} & m_{2} & m_{3}\end{array}\right)\left(\begin{array}{ccc}
l_{1} & l_{2} & l_{3}^{\prime}\\
m_{1} & m_{2} & m_{3}^{\prime}\end{array}\right) & = &
\frac{\delta_{l_{3}l_{3}^{\prime}}\delta_{m_{3}m_{3}^{\prime}}}{2l_{3}+1}\\
\sum_{l_{1}=|l_{2}-l_{3}|}^{l_{2}+l_{3}}\sum_{m_{1}=-l_{1}}^{l_{1}}
(2l+1)\left(\begin{array}{ccc}
l_{1} & l_{2} & l_{3}\\
m_{1} & m_{2} & m_{3}\end{array}\right)\left(\begin{array}{ccc}
l_{1} & l_{2} & l_{4}\\
m_{1} & m_{2}^{\prime} & m_{3}^{\prime}\end{array}\right) & = &
\delta_{m_{2}m_{2}^{\prime}}\delta_{m_{3}m_{3}^{\prime}}\\
\sum_{m=-l}^{l}(-1)^{l-m}\left(\begin{array}{ccc}
l & l & \ell\\
m & -m & 0\end{array}\right) & = & \sqrt{2l+1}\delta_{\ell,0}\,.
\end{eqnarray*}
The last expression is particularly useful in the derivation of
(\ref{SI}). 
\end{itemize}

\subsection*{Derivation of (\ref{Clmell})}

We start by equating expressions (\ref{fc-aniso}) and (\ref{APS})
\begin{equation}
\sum_{\ell}\sum_{l,m}\frac{2\ell+1}{\sqrt{4\pi}}\,\mathcal{C}_{\ell}^{lm}\,
P_{\ell}(\cos\vartheta)
Y_{lm}(\hat{n})=\sum_{\ell_{1},m_{1}}\sum_{\ell_{2},m_{2}}\langle 
a_{\ell_{1}m_{1}}a_{\ell_{2}m_{2}}^{*}\rangle
Y_{\ell_{1}m_{1}}(\hat{n}_1)Y_{\ell_{2}m_{2}}(\hat{n}_2)\,.
\label{CLlm-alm}
\end{equation}
As mentioned in the main text, the inversion of $\mathcal{C}_{\ell}^{lm}$
as a function of the $a_{\ell m}$'s is not a trivial task, since
the angles $(\Theta,\Phi,\theta)$ depend non-linearly on the angles
$(\theta_1,\varphi_1,\theta_2,\varphi_2)$. The easiest way to achieve this
goal is to pick up a coordinate system where only the $\theta$
dependence is present. After integrating it out, we rotate our coordinate
system using three Euler angles to recover back the $(\Theta,\Phi)$
dependence, which can then be integrated with the help of some Wigner
matrices identities. We start by positioning the vectors
$\hat{n}_{1}$ and $\hat{n}_{2}$ in the $xy$ plane, i.e, we chose
$\hat{n}_{1}=(\pi/2,\phi_{1})$, $\hat{n}_{2}=(\pi/2,\phi_{2})$. By
(\ref{cos-theta}) we then have $\cos\vartheta=\cos(\phi_{1}-\phi_{2})$.
Using the relation \cite{Varshalovich:1988ye}
\begin{equation}
Y_{\ell m}(\pi/2,\phi)=\lambda_{\ell
m}e^{im\phi},\quad\lambda_{\ell m}
=\begin{cases}
(-1)^{\frac{\ell+m}{2}}\sqrt{\frac{2\ell+1}{4\pi}\frac{(\ell+m-1)!!}
{(\ell+m)!!}\frac{(\ell-m-1)!!}{(\ell-m)!!}}\quad & 
\mbox{if}\quad\ell+m\in2\mathbb{N}\\
0 & \mbox{otherwise}
\end{cases}
\label{lambda-lm}
\end{equation}
we can integrate the $\theta$ dependence on both sides of (\ref{CLlm-alm}).
This gives us 
\begin{equation}
\frac{1}{\sqrt{\pi}}\sum_{l,m}\mathcal{C}_{\ell}^{lm}Y_{lm}(0,0)=\sum_{
\ell_{1},m_{1}}\sum_{\ell_{2},m_{2}}\langle 
a_{\ell_{1}m_{1}}a_{\ell_{2}m_{2}}^{*}\rangle
I_{\ell_{1}m_{1}\ell_{2}m_{2}}^{\ell}
\label{CLlm-alm2}
\end{equation}
where we have introduced the following definition 
\begin{equation}
I_{\ell_{1}m_{1}\ell_{2}m_{2}}^{\ell}\equiv-\lambda_{\ell_{1}m_{1}}
\lambda_{\ell_{2}m_{2}}
\int_{0}^{\pi}P_{\ell}(\cos(\varphi_{1}-\varphi_{2}))e^{i(m_{1}\varphi_{1}-
m_{2}\varphi_{2})}d(\cos(\varphi_{1}-\varphi_{2}))\,.
\label{I-ell-l1-l2-m}
\end{equation}

We need now to integrate out the $\Theta$ and $\Phi$ dependence
in the right-hand side of (\ref{CLlm-alm2}) which was hidden due
to our choice of a particular coordinate system. In order to do that,
we keep the vectors $\hat{n}_1$ and $\hat{n}_2$ fixed and make a rotation
of our coordinate system using three Euler angles
$\omega=\{\alpha,\beta,\gamma\}$. This rotation changes the coefficients
$\mathcal{C}_{\ell}^{lm}$'s and $a_{\ell m}$'s according
to 
\[a_{\ell
m}=\sum_{m^{\prime}}D_{mm^{\prime}}^{\ell}(\omega)\widetilde{a}_{\ell
m^{\prime}}\,,
\qquad\mathcal{C}_{\ell}^{lm}=\sum_{m^{\prime}}D_{mm'}^{l}
(\omega)\widetilde{\mathcal{C}}_{\ell}^{lm'}\]
where $\widetilde{\mathcal{C}}_{\ell}^{lm}$ and $\widetilde{a}_{lm}^{}$
are the multipolar coefficients in the new coordinate system and where
$D_{mm'}^{l}(\omega)$ are the elements of the Wigner rotation matrix.
The advantage of positioning the vectors $\hat{n}_{1}$ and $\hat{n}_{2}$ in
the plane $xy$ is that now the angles $\Theta$ and $\Phi$ are given
precisely by the Euler angles $\beta$ and $\gamma$, regardless of the value
of
$\alpha$\[\sum_{l,m}\mathcal{C}_{\ell}^{lm}Y_{lm}(0,0)=\sum_{l,m'}
\widetilde{\mathcal {C}}_{\ell}^{lm'}
\left(\sum_{m}D_{mm'}^{l}(\alpha,\beta,\gamma)Y_{lm}(0,0)\right)=\sum_{l,m'
}\widetilde{\mathcal{C}}_{\ell}^{lm'}Y_{l,-m}(\beta,\gamma)\]
 where in the last step we have used
$Y_{lm}(0,0)=\sqrt{(2l+1)/4\pi}\,\delta_{m0}$.
Therefore, in our new coordinate system we have (dropping the {}``
$\widetilde{}$ '' in our notation) \[
\frac{1}{2\pi}\sum_{l,m}\mathcal{C}_{\ell}^{lm}D_{0m}^{l}(\omega)\sqrt{2l+1
}=\sum_{\ell_{1},m_{1}}\sum_{\ell_{2},m_{2}}\langle
a_{\ell_{1}m_{1}}a_{\ell_{2}m_{2}}^{*}\rangle\sum_{m_{1}^{\prime}m_{2}^{
\prime}}I_{\ell_{1}m_{1}^{\prime}\ell_{2}m_{2}^{\prime}}^{\ell}D_{m_{1}^{
\prime}m_{1}}^{\ell_{1}}(\omega)D_{m_{2}^{\prime}m_{2}}^{\ell_{2}*}
(\omega)\,.\]
 We may now isolate $\mathcal{C}_{\ell}^{lm}$ using the identities
\cite{Varshalovich:1988ye} 
\begin{eqnarray*}
\int d\omega\,
D_{m_{1}m_{1}^{\prime}}^{l_{1}*}(\omega)D_{m_{2}m_{2}^{\prime}}^{l_{2}}
(\omega) & = &
\frac{8\pi^{2}}{2l_{1}+1}\delta_{l_{1}l_{2}}\delta_{m_{1}m_{2}}\delta_{m_{1
}^{\prime}m_{2}^{\prime}}\\
\int d\omega\,
D_{m_{1}^{\prime}m_{1}}^{l_{1}}(\omega)D_{m_{2}^{\prime}m_{2}}^{l_{2}}
(\omega)D_{m_{3}^{\prime}m_{3}}^{l_{3}}(\omega) & = &
8\pi^{2}\left(\begin{array}{ccc}
l_{1} & l_{2} & l_{3}\\
m_{1}^{\prime} & m_{2}^{\prime} &
m_{3}^{\prime}\end{array}\right)\left(\begin{array}{ccc}
l_{1} & l_{2} & l_{3}\\
m_{1} & m_{2} & m_{3}\end{array}\right)
\end{eqnarray*}
where $d\omega=\sin\beta d\beta d\alpha d\gamma$, to obtain 
\begin{eqnarray*}
\frac{1}{\sqrt{2l+1}}\mathcal{C}_{\ell}^{lm}&=&2\pi\sum_{\ell_{1},m_{1}}
\sum_{\ell_{2},m_{2}}\langle
a_{\ell_{1}m_{1}}a_{\ell_{2}m_{2}}^{*}\rangle\sum_{m_{1}^{\prime}m_{2}^{
\prime}}I_{\ell_{1}m_{1}^{\prime}
\ell_{2}m_{2}^{\prime}}^{\ell}(-1)^{m_{2}+m_{2}^{\prime}+m} \\ & &
\qquad\qquad\qquad\qquad
\left(\begin{array}{ccc}\ell_{1} & \ell_{2} & l\\ 
m_{1}^{\prime} & -m_{2}^{\prime} & 0\end{array}\right)\! 
\left(\begin{array}{ccc} \ell_{1} & \ell_{2} & l\\
m_{1} & -m_{2} & -m\end{array}\right)\,.
\end{eqnarray*}
If we now redefine $-m_2\rightarrow m_2$  and note that the first 3-j symbol above is 
identically zero unless $m'_{1}=m'_{2}$, we finally obtain (\ref{Clmell}).

\subsection*{Some properties of the integral (\ref{Int-l-ell})}

The geometrical coefficients $I_{\ell_{1}\ell_{2}}^{l,\ell}$ defined
in (\ref{Int-l-ell}) has many interesting properties which can be
explored in order to speed up numerical computation of (\ref{Clmell}).
First, we note that it is symmetric under permutation of $\ell_{1}$
and $\ell_{2}$:
\begin{eqnarray*}
I_{\ell_{1}\ell_{2}}^{l,\ell} & = &
\sum_{m}I_{\ell_{1}m\ell_{2}m}^{\ell}(-1)^{m}\left(\begin{array}{ccc}
\ell_{1} & \ell_{2} & l\\
m & -m & 0\end{array}\right)\\
 & = &
\sum_{m}I_{\ell_{2}m\ell_{1}m}^{\ell}(-1)^{m+\ell_{1}+\ell_{2}+l}
\left(\begin{array}{ccc}
\ell_{2} & \ell_{1} & l\\
-m & m & 0\end{array}\right)\\
 & = &
\sum_{m}I_{\ell_{2}m\ell_{1}m}^{\ell}(-1)^{m+2(\ell_{1}+\ell_{2}+l)}
\left(\begin{array}{ccc}
\ell_{2} & \ell_{1} & l\\
m & -m & 0\end{array}\right)\\
 & = & I_{\ell_{2}\ell_{1}}^{l,\ell}\,.
\end{eqnarray*}
Some of the other properties are a consequence of the integral
$I_{\ell_{1}m\ell_{2}m}^{\ell}$
defined in (\ref{I-ell-l1-l2-m}). We may note for example that, due
to the symmetry of the $\lambda_{\ell m}$ coefficient defined in
(\ref{lambda-lm}), we will have:
\[
I_{\ell_{1}\ell_{2}}^{l,\ell}=0\,,\quad\mbox{for any}
\quad\{(\ell_{1},\ell_{2})\in\mathbb{N}\,|\,\ell_{1}+\ell_{2}
=\mbox{odd}\}\,.
\]
Furthermore, the $\lambda_{\ell m}$ coefficients restrict the $m$
summation above to their values which obey:
$m+\ell_{1}+\ell_{2}=\mbox{even}$.
If we further notice that (\ref{I-ell-l1-l2-m}) is proportional to
the integral of a integral of the form 
$\int_{-1}^{1}P_{\ell}(\cos\theta)\cos m\theta\, d\theta$,
and that this integral is zero unless $\ell+m=\mbox{even}$, we conclude
that
\[I_{\ell_{1}\ell_{2}}^{l,\ell}=0\,,\quad\mbox{for
any}\quad\{(\ell_{1},\ell_{2},\ell)\in\mathbb{N}\,|\,\ell_{1}+\ell_{2}
+\ell=\mbox{odd}\}\,.
\]
Besides, using the fact that the integral
$\int_{-1}^{1}P_{\ell}(\cos\theta)\cos m\theta\, d\theta$
is zero for any $m<\ell$, we find
\[
I_{\ell_{1}\ell_{2}}^{l,\ell}=0\,,\quad\mbox{for
any}\quad\{(\ell_{1},\ell_{2},\ell)\in\mathbb{N}\,|\,\ell_{1}<\ell,\ell_{2}
<\ell\}.
\]
We finally comment on the special case where $l=0$, for which we
have
\[
I_{\ell_{1}\ell_{2}}^{0,\ell}=\frac{(-1)^{\ell_{1}}}{\sqrt{2\ell_{1}+1}}
\left(\sum_{m}I_{\ell_{1}m\ell_{1}m}^{\ell}\right)\delta_{\ell_{1}\ell_{2}}
\,.
\]
However 
\begin{eqnarray*}
\sum_{m=-\ell'}^{\ell'}I_{\ell'm\ell'm}^{\ell} & = &
\int_{0}^{\pi}P_{\ell}(\cos\vartheta)\left(\sum_{m=-\ell'}^{\ell'}\frac{
(2\ell'+1)}{4\pi}\frac{(\ell'+m-1)!!}{(\ell'+m)!!}\frac{(\ell'-m-1)!!}{
(\ell'-m)!!}e^{im\vartheta}\right)d(-\cos\vartheta)\\
 & = & \frac{2\ell'+1}{4\pi}\int_{-1}^{1}P_{\ell}(x)P_{\ell'}(x)dx\\
 & = & \frac{1}{2\pi}\delta_{\ell\ell'}\,.
\end{eqnarray*}
where in the derivation above we have made use of the Fourier series
expansion of the Legendre polynomial. So we conclude that 
\begin{equation}
I_{\ell_{1}\ell_{2}}^{0,\ell}=\frac{(-1)^{\ell_{1}}}{2\pi\sqrt{2\ell_{1}+1}
}
\delta_{\ell\ell_{1}}\delta_{\ell_{1}\ell_{2}},\label{IL-zero}
\end{equation}
which is needed in the derivation of (\ref{SI}).

\begin{eqnarray*}
\end{eqnarray*}

\end{document}